%% file: ms.tex
\documentclass[12pt,preprint]{aastex}
\shorttitle{FIR photometry of late-type Virgo Cluster galaxies}
\shortauthors{Tuffs et al. }
\usepackage{graphicx}
\usepackage{subfigure}
\usepackage{lscape}

\begin{document}

\title{Far-Infrared photometry of a statistical sample\\ of late-type Virgo 
Cluster galaxies
\footnote{Based on observations with ISO, an ESA project with 
instruments funded by ESA member states (especially the PI countries:
France, Germany, the Netherlands and the United Kingdom) and with the
participation of ISAS and NASA.} }

\author{Richard J. Tuffs,
Cristina C. Popescu\altaffilmark{2,3,4},
Daniele Pierini\altaffilmark{5},  Heinrich J. V\"olk}
\affil{Max Planck Institut f\"ur Kernphysik, Saupfercheckweg 1, 
69117 Heidelberg, Germany}

\author{Hans Hippelein}
\affil{Max Planck Institut f\"ur Astronomie, K\"onigstuhl 17, 69117 Heidelberg, Germany}

\author{Kieron Leech, Leo Metcalfe}
\affil{ISO Data Centre, Astrophysics Division, Space Science Department of
   ESA, Villafranca del Castillo, P.O. Box 50727, 28080 Madrid, Spain}
\author{Ingolf Heinrichsen and Cong Xu}
\affil{IPAC(Caltech/JPL), 770 S. Wilson Avenue, Pasadena, California 91125,
USA}
\altaffiltext{2}{present address: The Observatories of the Carnegie Institution of Washington,
813 Santa Barbara Str., Pasadena, 91101 California, USA; 
email:popescu@ociw.edu}
\altaffiltext{3}{Otto-Hahn Fellow of the Max Planck Institut f\"ur 
Astronomie, K\"onigstuhl 17, 69117 Heidelberg, Germany}
\altaffiltext{4}{Research Associate, The Astronomical Institute of the 
Romanian Academy, Str. Cu\c titul de Argint 5, Bucharest, Romania}
\altaffiltext{5}{present affiliation: The University of Toledo, Toledo, Ohio 43606-3390, USA} 

\begin{abstract}
We present deep diffraction-limited far-infrared (FIR) strip maps
of a sample of 63 galaxies later than S0 and brighter than $B_{\rm T}$ 16.8,
selected from the Virgo Cluster Catalogue of 
Binggeli, Sandage \& Tammann (1985). 
The ISOPHOT instrument on board the Infrared Space Observatory was 
used to achieve sensitivities typically an order of 
magnitude deeper than IRAS in the 60 and 100\,${\mu}$m bands and to 
reach the 
confusion limit at 170\,${\mu}$m. 
%This latter wavelength was previously largely
%unexplored, as IRAS had no spectral coverage longwards of 100\,${\mu}$m. 
The averaged 3\,$\sigma$ upper limits for integrated flux densities of 
point sources at 60, 100 and 170\,${\mu}$m are 43, 33 and 58\,mJy, 
respectively. 
%The faintest 3\,$\sigma$ upper limits
%at the three wavelengths are 30, 20 and 40\,mJy. 
54 galaxies (85.7$\%$)
are detected at least at one wavelength, and 40 galaxies 
(63.5$\%$) are detected at all three wavelengths. 
The highest
detection rate (85.7$\%$) is in the 170\,${\mu}$m band.
In many cases the galaxies are resolved, allowing the scale length of the 
infrared disks to be derived from the oversampled brightness profiles 
in addition to the spatially integrated emission.
The data presented should provide the basis for a variety of
statistical investigations of the FIR spectral energy distributions of gas 
rich galaxies in the local universe spanning a broad range in star-formation 
activity and morphological types, including dwarf systems
and galaxies with rather quiescent star formation activity.

\end{abstract}


\section{Introduction}

Galaxy emission in the  Far Infrared (FIR) is intimately connected to the 
current rate of star formation, since the
characteristic radiation of young stars in the ultraviolet and optical
wavelength range will at least in part be absorbed by interstellar dust. 
The IRAS all sky survey - 
within its limitations to wavelengths shorter than about 120\,${\mu}$m - 
raised statistical studies of star formation in galaxies to a new level, 
detecting over 25,000 objects (see Soifer et al. 1987 
for a review). Despite the bright (by optical standards)
detection limit of $\sim\,0.5$ and 1.5\,Jy at 60 and 100\,${\mu}$m, 
respectively, about half of the IRAS galaxies had no counterparts
in optical catalogues at the time of the survey. In part, this was 
a consequence of the almost complete sky coverage and homogeneous
data processing of the IRAS mission. However, it also reflected the 
fact that at luminosities greater than about $10^{11}\,{\rm L}_{\odot}$
optically selected galaxies are less common than IRAS galaxies. These
luminous infrared ojects are thought to be relatively distant systems
undergoing intense bursts of massive star formation, probably triggered by 
mergers and interactions, in which the bulk of the stellar luminosity is 
locally absorbed by dust and re-radiated in the FIR 
(see e.g. Sanders \& Mirabel 1996 for a review, Leech et al. 1994). 

In the local Universe the most common gas rich systems are the 
so-called ``normal'' late-type galaxies (later than S0). By ``normal'' we 
loosely refer to galaxies which are not dominated by an active nucleus and 
whose current star 
formation rates (SFRs) would be sustainable for a substantial fraction 
of a Hubble time. These objects are intrinsically important, since, apart 
from the diffuse intergalactic gas, they comprise
the dominant fraction of the baryonic matter in the universe. Perhaps
of even more interest for structure formation and extragalactic 
astronomy is the fact that normal late-type galaxies are the best objects 
for the investigation of the still poorly understood mechanisms for
star formation in galaxian disks, as well as the associated 
global physical processes and their interrelations. 

The deepest statistical investigation of normal galaxies in the local Universe
using IRAS was that of Devereux \& Hameed (1997) who, 
after processing the data with the latest techniques, 
examined the FIR luminosity function at 60 micron for 1215 galaxies 
within a distance of 40~Mpc, selected from the Nearby Galaxies Catalog 
of Tully (1989). Even in this catalogue low optical luminosity galaxies and 
in particular low optical surface brightness galaxies are prone to be 
under-represented. Furthermore, as discussed
by Devereux \& Hameed, the sensitivity limit of the IRAS survey
meant that only limited information could be derived for the 
FIR properties of less IR luminous objects and, in particular,
of galaxies with almost quiescent star formation activity.

%Thus, although the 
%detection rate for spirals of type Sb-Sbc at 60 ${\mu}$m was almost 
%100 percent out to a distance of about 15~Mpc, the detection rate for Im
%galaxies was only 15 percent even for the most nearby objects 
% (those within 3~Mpc).  

Further obvious biases of IRAS studies in general
are the lack of spectral coverage longwards of the
100\,${\mu}$m filter (for which the FWHM system response
is approximately 80 - 120\,${\mu}$m - see Fig. II-C-9 of Beichman et al.
1988) and the three times brighter sensitivity
limit in this band compared to the IRAS 60\,${\mu}$m band. This could translate
into a bias against the detection of FIR emission from quiescent systems, 
if the 60/100\,${\mu}$m colour ratio is indeed correlated with 
massive star formation activity as proposed by e.g. Lonsdale \& Helou 1987. 
More fundamentally, the suggestion of Chini et al. (1986)
that the spatially integrated spectral energy distribution (SED) of late--type
galaxies peaks in the 100 - 200\,${\mu}$m range and cannot be simply
extrapolated by fitting the IRAS 60 and 100\,${\mu}$m photometric points
with single temperature dust emission components is now confirmed
by initial studies with ISO (e.g. Stickel et al. 2000; 
Popescu et al. 2002).

By virtue of its superior intrinsic sensitivity, and the 
availability of longer integration times than were possible with IRAS,  
the ISOPHOT instrument (Lemke et al. 1996) on board ISO 
(Kessler et al. 1996) could detect discrete sources at 
least 10 times fainter than IRAS at 60 and 100\,${\mu}$m. It furthermore
had a wavelength coverage extending to 240\,${\mu}$m.
The basic observational goal of the project we present here is to use ISO
to extend knowledge of the FIR SEDs to lower 
luminosity limits 
and to cover the peak in ${\nu}$S$_{\nu}$ for a complete sample of 
normal late type galaxies, embracing a large range in morphological 
type and star formation activity.

Ideally, one would seek to achieve
this by making a blind survey in the FIR by mapping a substantial 
fraction of the sky, analogous to IRAS. However, 
the optimum sensitivity to luminosity of
a detection survey with ISOPHOT was for targets at such a distance that 
the angular size of the FIR emission region matched the
angular resolution of ISO (45 arcsec at 100 micron). 
For typical dwarf galaxies this corresponds to a distance of 
about 15\,Mpc, so that the low surface number density of such close-by objects 
would have made a blind survey too costly. 

An almost ideal alternative basis for a blind survey in the FIR is 
offered by the Virgo Cluster Catalogue (VCC), obtained by
Binggeli, Sandage \& Tammann (1985; hereafter BST85) in their deep
blue photographic survey. We therefore selected a sample of late-type
galaxies from the VCC catalogue to be observed with ISOPHOT.
From an observational point of view the Virgo cluster has the advantage that it
is situated at high galactic latitude and is close to the ideal distance for
the detection of dwarf galaxies with ISOPHOT.
The VCC has a full
representation of morphological types of normal gas rich galaxies, 
including  quiescent systems and even, to some extent, low surface 
brightness objects, ranging from bright ($B_{\rm T}\,\sim\,10$) 
giant spirals down to blue compact dwarfs (BCDs) and irregular galaxies 
at the completeness level of $B_{\rm T}\,\sim\,18$. 
% ($M_{BT}\,\sim\,$-13.7). 
Moreover the VCC galaxies are the most carefully classified in
terms of optical morphology.
Virgo cluster galaxies have been also extensively studied in the UV (Deharveng
et al. 1994), optical (e.g. Schr\"oder 1995), $H_{\alpha}$, NIR 
(Boselli et al. 1997a),
radio continuum (Gavazzi \& Boselli 1999, Niklas, Klein \& Wielebinski 1995), 
HI (Hoffmann et al 1989 \& references therein) and CO (e.g. Boselli, Casoli \&
Lequeux 1995).

From an astrophysical point of view the cluster is ideal in that it is known 
to be a dynamically young system, with a significant fraction of galaxies 
freshly falling in from the
field (Tully \& Shaya 1984, Binggeli, Popescu \& Tammann 1993). The 
fundamental incentive for
choosing the VCC as the basis of a statistical sample for ISOPHOT
was thus that a luminosity- {\it and} volume - limited sample
of cluster periphery and cluster core galaxies representative of the field and
cluster environments, respectively, could be observed down to the least 
luminous dwarf galaxies reachable with ISOPHOT. This should
allow an investigation of the strength and time-dependence of all 
manifestations of star formation activity and its relation to intrinsic 
galaxy properties such as Hubble type or sheer overall size.
Clusters are natural 
laboratories for the investigation of the effect on galaxy 
properties of external conditions such as the pressure and mass density of the 
diffuse intracluster medium (ICM), the large-scale gravitational potential 
of the cluster, or the interaction with other galaxies. This is
particularly the case for the Virgo cluster for which deep X-ray
maps (B\"ohringer et al 1994) of the ICM, as well as comprehensive
measurements of the HI deficiences (also for dwarf galaxies) are
available (Hofmann et al. 1989 and references therein). Another potential
manifestation of galaxy-ICM interaction would be the detection of FIR emission 
in the vicinity of infalling spirals, arising from collisionally heated grains
released from the disk through ram-pressure stripping (Dwek, Rephaeli \& Mather
1990, Popescu et al. 2000a).

%One of the main statistical challenges to be expected in intercomparisons
%between cluster core and cluster periphery galaxies, 
%and between the ISOPHOT Virgo sample and field samples is the
%unravelling of the type-density relation 
% (e.g. Dressler 1984), especially in cases where the FIR
%characteristics depend on both environment and morphological type. 
%This is essentially the price to be paid for the
%completeness and depth offered by a cluster-based sample. 
To maximise the statistics attainable in the fixed available observation 
time of 20.36 hours, the survey was made in just three broad band-passes, 
centred at 60, 100 and 170\,${\mu}$m, which together encompass 
the range 40 - 240\,${\mu}$m. In all, a subset of 63 galaxies 
with $B_{\rm T}\,\le\,16.8$ from the VCC were observed. These were selected 
from specific regions of the sky as described in Sect.~2.

A basic aim of the survey was to derive integrated properties of the
sample. This is especially appropriate for cluster samples at a fixed 
distance to enable the construction of FIR luminosity functions. 
Furthermore, the angular resolution of ISO
(at 100\,${\mu}$m - 4 times superior to that of IRAS)  
was sufficient to resolve the disks of the giant spirals. This allowed
a subsidiary science goal to be realised - namely the provision
of statistical information on scale lengths of spiral galaxies in the 
FIR, their relation to the optical scale lengths, and, to a limited 
extent, the separation of nuclear and disk emission components. 
An oversampled scanning technique was used to achieve
the combined goals of obtaining integrated fluxes, structural information,
and robustness against confusion with background sources or foreground cirrus 
emission, which limits the survey at the longest wavelength.
The use of long scans means that our survey is sensitive
to the emission of cold grains in the outer regions of the disks, which are 
prone to be missed or under-represented in pointed observations of 
galaxies. 

The combination of spatially integrated photometry extending in wavelength to
170\,${\mu}$m will provide the basis for the modelling of the  SED of normal 
galaxies.
%Unlike the optically thick distant luminous systems, normal galaxies in 
%the local universe typically have intermediate optical depths to starlight
% (though whether the bias is to the optically thin or thick regime is
%still a subject of vigorous debate). Apart from the fact that 
%galaxy disks are inhomogeneous, and also harbour luminous 
%optically thick components, the relevance of the FIR to 
%studies of star formation here is that 
In this wavelength range  grains act as test particles 
probing the strength and colour of the diffuse UV/optical radiation fields. 
This constitutes an entirely complementary constraint to direct  
studies of stellar emission in the UV and optical bands. The SED modelling
should yield statistical information on the disk opacities and the 
relative contributions of young and old stellar populations to the dust 
heating (see e.g. Xu \& Buat 1995; Silva et al. 1998; 
Devriendt, Guiderdoni \& Sadat 1999; Popescu et al. 2000b, 
Misiriotis et al. 2001). The star 
formation rate and star formation histories can in principle be derived from 
such techniques. 

Another scientific goal of the survey is to study the statistical correlation
of FIR emission of normal disk galaxies with other emissions from the radio 
to the X-ray spectral regions. In particular, an
unexpected result from the IRAS mission was the especially tight correlation 
between the radio 
synchrotron and the FIR emission
(e.g. Helou, Soifer \& Rowan-Robinson 1985, de Jong et al. 1985, 
Wunderlich, Wielebinski \& Klein 1987). The link
is qualitatively assumed to be given by the grain heating associated with
the appearence of massive stars and the acceleration of relativistic
particles in their eventual SN explosions. Basically the correlation can be
explained in term of a calorimetric theory (V\"olk 1989, 
Lisenfeld, V\"olk \& Xu 1996). 
One question raised by the IRAS studies is whether the effects of a 
cluster environment will indeed shift the FIR/radio - correlation 
on average to considerably higher radio emission for given FIR 
intensity as claimed by Gavazzi, Boselli \& Kennicutt (1991). 
%This might indicate 
%the existence of interaction processes which increases energetic particle 
%production in galaxies without changing the FIR properties (V\"olk \& Xu 1994).
Another issue is the extent to which the constraints imposed on galaxian
properties by the correlation will apply to less massive and/or more 
quiescent systems than those typically studied with IRAS.

The sample of VCC galaxies selected for ISOPHOT also formed a substantial part 
of samples observed with ISO using the ISOCAM MIR camera in pass bands 
centred at 6.9 and 15\,${\mu}$m (Boselli et al. 1997b, 1998) and the LWS at the
158\,$\mu$m [CII] fine structure gas cooling line 
(Leech et al. 1999).
In particular, the latter observations provide complementary information
about the energetics of the interstellar gas, the sources of [CII] emission
within the interstellar medium (Pierini et al. 1999, 2001a), and
the role played by different stellar populations in the gas heating
(Pierini et al. 2001b).

%These studies, partly by themselves, 
%and in combination with others, have already led to a number 
%of interesting results, like the equal
%contribution to the $158 \mu$ flux from dense photo-dissociation regions, 
%and from the diffuse neutral interstellar gas, as well as the existence 
%of "[CII]-quiet" galaxies that have a star formation rate below a 
%threshold not reachable by either the
%IRAS instrument or the FIR spectrometer on the Kuiper Airbone Observatory.

% and with the LWS in the [CII] 158~${\mu}$m
% cooling line (Pierini et al.,   ***).

This paper is organised as follows. 
In Sect. 2 we describe the selection criteria,
optical properties and completeness of the observed sample.
Observational details and a description of the data analysis are given in
Sections 3 and 4, respectively. The photometric characteristics of the survey, 
including comparisons with IRAS and COBE photometry, are given in Sect. 5.
The extraction of the integrated photometry and the resulting catalogue with 
associated plots are presented in Sect.~6. Basic properties
and detection statistics are derived in Sect.~7 and the main results of the
paper are summarised in Sect.~8.

\section{The sample}

A sample of 62 VCC galaxies selected from the Virgo Cluster Catalogue
(BST85) with Hubble type later than S0 was selected according to criteria 
based on their projected position and magnitude. 

The sample galaxies lie in two sky areas called ``core'' and ``periphery'',
where the spread in distance within the sample, due to the complex
3D structure of the cluster (Binggeli, Popescu \& Tammann 1993 and 
references therein), is minimised.
These areas were chosen in order to optimise the statistical evaluation
of the environmental effects of the cluster, if any, on the observed 
properties of galaxies of the same morphology.
It is well-known that galaxies in the cluster core are HI-deficient
(Haynes \& Giovanelli 1984) while galaxies in the cluster periphery
have the same HI content as field galaxies.

The so-called {\it cluster core subsample} is comprised of 25 certain
spiral galaxies brighter than $B_{\rm T} = 14.5$, with 
projected positions within 2 degrees of M~87 (Fig.~1), essentially seen 
towards the extended X-ray halo
of M~87 (B\"ohringer et al. 1994). These galaxies define an optically 
complete sample of galaxies later than S0 and earlier than Im. 
{\it The cluster periphery subsample} is comprised of 37 spiral,
irregular or BCD galaxies with $B_{\rm T} \leq 16.8$ and 
with angular separations of greater than 4 degrees from the position
of maximum projected galaxy density given by Binggeli et al. (1987) (Fig.~1).
Objects with RA(1950.0) $\rm > 12.596^{\rm h}$
or within 1.5 degree of the position of maximum projected galaxy 
density of M~49 sub-cluster are excluded. In addition objects seen towards the
M and W clouds (as defined in Fig.~1 of Sandage et al (1985) and also
reproduced in our Fig.~1) or in the Southern extension 
($\delta \rm < 5^o$) are also excluded.
The limiting magnitude of $B_{\rm T}=16.8$ is well within the VCC 
completeness limit of $B_{\rm T}\,\sim\,18$.
These two subsamples constitute two volume-and magnitude-limited samples
of late--type galaxies. The selected galaxies consist of 60 members and 2
possible members (VCC~130 and VCC~890) of the Virgo cluster (Binggeli, Popescu
\& Tammann).

Due to technical problems with spacecraft operations for some observations, 
60 out of the 62 selected VCC galaxies were actually observed
with ISOPHOT, giving a statistical completeness of 97\%.
In particular, all the 25 galaxies of the core subsample and 35 out of
the 37 galaxies of the periphery subsample were observed, the two observed
subsamples having a statistical completeness of 100\% and of 94\%,
respectively.
Hereafter, we refer to the 60 VCC galaxies observed by us with ISOPHOT
as the {\it statistical sample} (marked as circles in Fig.~1).

In addition to this statistical sample, an extended program to observe
the remaining 20 cluster member late--type galaxies fainter than 
$B_{\rm T} = 14.5$ from the cluster core region was initiated to provide
a statistical comparison to faint galaxies (most particularly
the Im and BCD galaxies) in the cluster periphery subsample.
However, due to scheduling and visibility constraints, only 3 galaxies
(VCC~1001, VCC~1121 and VCC~1569) from this extended program
were actually observed (marked as crosses in Fig~1).

Table~1 lists the optical and FIR (IRAS) properties
of all the 63 VCC galaxies observed by us with ISOPHOT. 
Our statistical sample includes 7 early--type spiral galaxies
with claims of Seyfert/LINER activity or of both (cf. Tab. 1).
For VCC~836 (Philips \& Malin 1982), VCC~857 (Keel 1983), VCC~1690
and VCC~1727 (Stauffer 1982), these claims were raised before
the selection and observation of our statistical sample.
On the other hand, VCC~92 (Barth et al. 1998), VCC~1690 and VCC~1727
(Ho, Filippenko \& Sargent 1997) have recently been defined as transition
spirals, while VCC~857 (Rauscher 1995) and VCC~1043
(Ho, Filippenko \& Sargent 1997) seem to be marginal candidates
for non-stellar nuclear activity.
The phenomenology of Virgo cluster spiral galaxy nuclear regions has not
yet been established (e.g. Rauscher 1995), since different types of LINERs
(i.e., photoionized by a stellar continuum or by an active galactic
nucleus) cannot easily be distinguished from one another (e.g.
Alonso-Herrero, Rieke \& Rieke 2000). Given this, we still consider it
reasonable to include these 7 VCC spiral galaxies as part of our
statistical sample of normal late-type galaxies.

Fig.~2 shows the distribution of the objects of the core
and of the periphery subsamples (panels a and b, respectively)
in the plane defined by their Hubble type and total apparent
blue magnitude $B_{\rm T}$.
%In each panel, the last morphological bin contains either peculiar galaxies
%or objects of uncertain morphology. 
%Circles and squares denote objects in the core and in the periphery 
%subsamples, respectively.
The figure shows that all galaxies in the core subsample
and all periphery galaxies brighter than $B_{\rm T} = 16.5$
were observed (filled symbols). The core and periphery subsamples are thus
complete to $B_{\rm T} = 14.5$ and to $B_{\rm T} = 16.5$, respectively.
It is also evident that the periphery subsample is dominated
by spiral galaxies later than Sbc and by irregular or dwarf spiral galaxies,
while the core subsample is dominated by spiral galaxies of type Sc
and earlier. This behaviour reflects
the combination of the morphology--density relation with
the morphology--luminosity relation (Sandage, Binggeli \& Tammann 1985). In 
this respect, 
the two subsamples are complementary, so that our statistical sample 
represents the overall phenomenology
of the {\it normal} spiral, irregular or dwarf spiral galaxies brighter
than $B_{\rm T} = 16.5$.

\section{Observations}

All observations were made using ISOPHOT's ``P32'' Astronomical Observing 
Template (Tuffs $\&$ Gabriel 2001), which combines the 
standard spacecraft raster pointing mode (stepping in spacecraft 
Y and Z coordinates) with scans in Y made with the 
focal plane chopper. 
At each spacecraft raster position the 
focal plane chopper is stepped at intervals 
of one third of the detector pixel pitch, resulting
in sky samplings in Y of $\sim\,$15 and 31$\,^{\prime\prime}$ 
for the C100 and C200 detectors, respectively. This is close to the 
limit for Nyquist sampling of 
(17$\,^{\prime\prime}\times\lambda$)/100$\,{\mu}$m. 

The dimension of the spacecraft raster in Y was always made 
large enough to ensure that the scans spanned the 
projection of the optical diameters 
(to the 25.5\,mag\,arcsec$^{-2}$ 
B-band isophote) in Y, plus at least one independent resolution element
beyond this limit at each scan end.
This technique ensured that the survey was sensitive to any infrared emission 
from a galaxy with as large an extent as the optical extent, 
while at the same time sampling the target finely
enough to probe the basic morphology of the disk emission. 
These attributes were especially relevant to observations of the bright 
giant spirals in the sample. A particular aim was to be sensitive to 
any faint emission from cold dust in the outer disk.
Typically, only one spacecraft scan leg was made for the giant 
spirals. The only exception was the galaxy pair VCC1673/VCC1676, which
was mapped in two dimensions, to facilitate the morphological separation
of the two systems.
The single scan observations resulted in strip maps of dimension 3 and 2 
in the spacecraft Z coordinate for the C100 and C200 detector, respectively, 
separated by 46 and 92$^{\prime\prime}$.  Although undersampled in Z, the 
integrated properties of 
the galaxies could in almost all instances be recovered using the modelling 
procedures described in Sect.~6. In a few unfavourable cases where 
the scan direction in spacecraft Y approached being perpendicular to 
the major axis of the galaxy, some extended emission from the disk was 
missed.

The oversampling in Y afforded by the P32 mapping technique was 
also well suited to the fainter targets (generally BCD and irregular 
galaxies) in the sample, as it mitigated the effects of
confusion with the forground cirrus 
emission from the Milky Way. To reach the confusion limit integration times 
were increased 
for optically faint targets by increasing the dimension of the spacecraft
raster in Z, stepping in multiples of the detector pixel pitch so as 
to ensure that one row of the detector was always scanning through 
the target. This also improved the precision with which the flat field 
could be determined. Another problem in the detection of faint sources 
is the level of residual glitches. Here again, the P32 technique was 
beneficial to the detection of faint sources as the inherent rapid sampling
and redundancy allowed a deeper deglitch, with an efficient rejection 
of the longer lived responsivity fluctuations which often follow 
glitch events.

All galaxies were observed in three broad band filters - the C60 and 
C100 filters on the C100 detector and the C160 filter (which actually 
has a reference wavelength of 170\,${\mu}$m) on the C200 detector. 
The C100 filter was chosen in preference to the somewhat more sensitive 
C90 filter as it provided a more direct comparison with the IRAS 
100\,${\mu}$m observations of the brighter galaxies in the sample. 
This was useful as a check of the calibration (see Sect.~5). 
%A comparison between the ISO and IRAS measurements can also give 
%information on any emission components on larger angular scales 
%than the coverage of the P32 scans.
The observations of some galaxies were repeated due to
operational problems or severe detector instabilities induced by
radiation hits. In such cases only the dataset of highest quality
was used.

61 of the 63 galaxies were observed in the period May\,-\,July 1996.
The remaining two galaxies were observed in November 1997.
A summary of the basic observational parameters for each target is given in 
Table~2. 
%The table is organized as follows:

%\begin{itemize}
%\item{\it Column 1:} VCC denomination;
%\item {\it Column 2:} Target Dedicated Time (TDT) identifier;
%\item {\it Column 3:} filter;
%\item {\it Column 4:} raster pointing pattern (in spacecraft Y and Z); 
%\item {\it Columns 5 and 6} : The coordinates $\alpha$(2000.0) and 
%$\delta$(2000.0) of the map center (derived from pointing data);
%\item {\it Column 7:} position angle (PA) of the spacecraft Y axis 
%(positive East from North);
%\item {\it Column 8:} on-target integration time ($\ T_{\rm int}$);
%%\item {\it Column 9:} notes. Here, a dagger marks galaxies observed with 
%%ISOPHOT whose observations, although succesful, were repeated due to a 
%%problem elsewhere in the concatenated chain. 
%\end{itemize}

%\input{datareduction.tex}

\section{Data Reduction}

Data was reduced using procedures specifically developed
for the analysis of data taken in the ``P32'' Astronomical
Observing Template. These procedures are fully described by 
Tuffs \& Gabriel (2001); here we only outline the major aspects
relevant to the present analysis of the Virgo cluster galaxy data. The most
important functionality of this new software is the correction of the 
transient response behaviour of the Ge:Ga photoconductor detectors of ISOPHOT. 
As we show below in the context of the present observations, failure 
to correct for this effect in data taken in the ``P32'' mode can give 
rise to serious signal losses and distortions in the derived brightness 
profiles through the galaxies. 

In this paper the data reduction was made outside of the standard PIA 
(ISOPHOT Interactive Analysis package; Gabriel et al. 1997 ) program.
However, most of the basic operations needed for a photometric calibration 
of ISOPHOT data - dark current subtraction, correction for integration ramp 
and signal non-linearity effects, the reset interval correction, and the 
conversion of engineering units in V/s into astronomical units 
(MJy\,sr$^{-1}$) - utilised subroutines and calibration datasets 
taken from the standard PIA package. The only exception was the correction
for vignetting (see Tuffs \& Gabriel 2001).
In this work, the calibration 
is based on the V8.1 release of PIA. There were three principle steps 
involved in the reduction of ``P32'' data of the Virgo galaxies: signal 
conditioning, transient correction and calculation of calibrated transient
corrected maps.

\subsection{Signal Conditioning}

The first step, which is a non-iterative process applied individually
to each detector pixel, is {\bf signal conditioning}. This aimed to 
provide a signal timeline in engineering units free of certain 
instrumental artifacts, but retaining the complete signature of the 
transient response of the detector pixel to the illumination history
at the full time resolution of the input data. At this stage,
the data was corrected for integration ramp non-linearity effects
and an orbit dependent dark current was removed. The signal non-linearity
correction was not applied prior to the responsivity drift correction.

Information on the actual pointing history of the spacecraft during
the observation was extracted and interpolated to the time of each
elementary data sample taken during the galaxy observation. This was 
used to define a rectangular grid of pointing directions on the sky 
covered by the occuring combinations of chopper and spacecraft pointing
positions. Here, the grid spacing was chosen such that the pointing 
directions were within typically one arcsec of the centre of each grid
pixel. As the grid pixels are $\sim\,$15 and 31 arcsec wide in the
spacraft Y coordinate, this removes any need for interpolation or gridding
functions when making maps of the sky brightness distribution.
We refer to this grid of sky directions as the ``P32 natural grid''.
An example of the actual pointing directions viewed in detector
pixel 1 in an observation with the C100 detector is given in Fig. 3.

The last part of the signal conditioning was the application of  
P32 specific deglitching procedures to remove spikes and 
longer-lived fluctuations in detector responsivity from the data time line.
The spike detector was operated individually on the residuals of the data
remaining on each chopper plateau after fitting a polynomial to preserve
the slower varying transient effects. For bright compact sources observed
with the C100 detector, there was evidence for spontaneous glitching when the 
chopper traversed the direction of peak brightness. Since laboratory 
investigations on Ge:Ga photoconductors (see Sclar 1984) have shown that 
this glitching should be preserved as signal, 
a brightness-dependent threshold for the spike detector was employed 
to prevent serious signal loss through this effect. Lastly, longer-lived
``tails'' in the responsivity of detector pixels, often seen following 
a spike, were identified and flagged by analysing the temporal development
of the average level of each chopper plateau viewing a particular sky 
direction on a given spacecraft pointing. This procedure, which utilises 
the redundancy afforded by the P32 technique, is the most crucial
step determining the instrumental sensitivity, which in the
case of the observations of the Virgo galaxies is limited by the degree
to which the stochastic undulations in detector responsivity can be removed.
In general, removal of these artifacts was virtually complete for the C200
detector, but could only partially be achieved for the C100 detector.

\subsection{Transient Correction}

Following signal conditioning, the second major step in the data reduction
is {\bf transient correction}. This is an iterative process to
determine the most likely sky brightness distribution giving rise to
the observed signal timeline. It is a non-linear optimisation problem 
with the values of sky brightness distribution at each point of the P32 
natural grid as variables.
Iterations are performed involving successive comparisons between the 
input data time line and a timeline predicted from the convolution
of a non-linear detector model with a trial illumination history derived 
from a trial sky brightness distribution.

The basis for the solution is the empirical model for the transient behaviour
of the ISOPHOT C100 and C200 photoconductor detectors described in 
Tuffs $\&$ Gabriel. This involves the superposition of two exponentials, 
each with illumination dependent time constants.
The model also incorporates illumination-dependent jump factors
and responsivity coefficients. 
Each detector pixel has its own constants. 
As discussed by Tuffs $\&$ Gabriel, this model adequately represents the 
detector response on 
timescales greater than a few seconds, but only approximately describes the
so-called ``hook response''. This leads to an over-correction in the
photometry (downwards or upwards respectively for decreasing or
increasing illumination steps) when the chopper dwell time is
shorter than a few seconds. This was the case for the
P32 observations of Virgo galaxies which in almost all
cases had chopper dwell times of 0.6~s. Because of the illumination
dependence of the detector time constants and jump factors, the effect
is more pronounced for decreasing than for increasing illumination steps.
For bright sources with a high contrast to the background this
induces an asymmetry and a narrowing of the derived brightness
profile through a point source compared to the nominal point spread
function.
%The solution tends to 
%undershoot the true solution when an increase in illumination is
%encountered, and overshoots for a decrease in illumination. 
Where this was a blatant effect in the drift corrected data timeline, data was 
censored out manually.

Fig.~4 shows an example of a brightness profile through the faint 
standard star HR1654 at 100\,${\mu}$m, for data processing with and without 
the responsivity correction.
%An example of a brightness profile through the faint standard star
%HR1654 at 100\,${\mu}$m is given in Fig.~4, processed after the 
%responsivity correction is given in Fig.~4. For comparison, a dotted 
%profile has been overlaid representing an identical data processing but 
%without the responsivity correction. 
The star has a flux density of 0.52\,Jy 
which would correspond to a medium brightness Virgo galaxy in the sample.
Some 95 percent of the flux density has been recovered by the transient
correction procedure. Without the correction some 30 percent of the 
integrated emission is missing and 50 percent of the peak response.
The local minimum near 60 arcsec in the Y offset is a typical
hook response artifact where the algorithm has overshot the true solution
after passing through the source. The actual corrections in integrated
flux densities for Virgo
galaxies depend on the source brightness, the source/background ratio, and
the dwell time on each pointing direction, so there is no fixed correction
factor for the photometry. As a rule of thumb, though, brighter more
compact galaxies have larger corrections than extended faint sources.

The final part of the transient correction procedure is a reassignment of
the uncertainties in the photometry. This is done by examining the scatter
in the solutions obtained for each chopper plateau, and is a measure of the
irreproducibility of the solution caused by imperfections in the detector
model and/or glitch-induced stochastic undulations in the detector
responsivity (which, like the white detector noise, are amplified by the 
transient correction procedure). This procedure yields a more representative 
measure of the
true random uncertainty in the maps than uncertainties derived from the 
original fits to the integration ramps.

\subsection{Calculation of calibrated transient corrected maps}

This last stage in the processing starts with the conversion of the
transient corrected data from V/s to MJy/sr. For the Virgo cluster
galaxies the conversion factors were found individually for each galaxy 
using the average of the responses for each detector
pixel to the fine calibration source exposures made before and after each 
raster in each filter. Thereafter a time dependent flat field was performed,
by examining the time dependence of the detector response to the background
directions viewed at the most negative and most positive offsets in Y
from the map centre. A flat sky background emission was assumed.
For maps with a single spacecraft raster leg, only a
linear time dependence could be fitted. The deeper maps with three
scan legs could be fitted with a cubic polynomial.

Maps were made from the transient corrected, calibrated and flat fielded
data through a weighted coaddition of the primitive data samples at the
full time resolution at each point of the P32 natural grid. 
Data taken
on slews was not used. No 
gridding function was employed, so that the full angular resolution 
inherent to the data was preserved. This also means that the measurements 
at different positions on the map grid are from entirely independent
data sets. In cases where there was full detector pixel redundancy in the
C100 array
(for faint galaxies with at least three scan legs), the coaddition of the
data was done first for individual pixels, producing a map cube of
9 maps. Over the region of the P32 natural grid where all 9 maps
had data, data from the extreme two detector pixels in the map cube 
was not used when combining data from all pixels. This acted as a further
filter for the glitch-induced undulations in the signal which is the
major instrumental source of error. However,  this artifact could only be
partially removed in this fashion. Many of the final maps appear to exhibit
spatially correlated noise at a level greater than predicted by a simple
error propagation analysis, which we attribute to residual glitch tails in 
the data enduring over a series of chopper positions.

The final step is the subtraction of the background. This was done by
subtracting a tilted plane obtained from a fit to the extremities of the map 
(external to the extent of the optical galaxy). In principle, this should
provide a better statistical measure of the source brightness near the map 
centre than subtracting individual baselines from each map row.

\section{Evaluation of calibration}

The data reduction procedures outlined in Sect.~4 produced
maps of Virgo cluster galaxies in surface brightness units calibrated on the 
ISO flux scale. 
When comparing ISO data with that from other observatories account should
be taken of possible systematic variations in the absolute
calibration scale.
This section compares the ISO photometry for the Virgo survey with
that from COBE-DIRBE and IRAS. This is done firstly
by statistically comparing the surface brightnesses of the backgrounds
measured by ISOPHOT-C with predictions 
for these backgrounds derived from the COBE-DIRBE archives. This provides
information about the calibration of surface brightness in each of the
ISOPHOT C60, C100 and C160 bands. Secondly, a statistical comparison of 
integrated flux densities of the target galaxies in the C60 and C100
bands with IRAS measurements in the IRAS 60 and 100\,${\mu}$m bands
is made. This provides information about the calibration of 
discrete sources. On the basis of the analysis presented below
we decided to calibrate the ISO data on the 
COBE-DIRBE flux scale. We give the procedure for the conversion of the 
ISO brightnesses to the COBE-DIRBE flux scale, as well as conversion factors 
from COBE-DIRBE to the ISO and IRAS flux scales appropriate for the 
observations of Virgo cluster galaxies in the P32 mode.

\subsection{Comparison of ISO and COBE-DIRBE Background Measurements}

Background brightnesses were derived for the 63 VCC galaxies observed by 
ISOPHOT
in each of the COBE-DIRBE bands. The procedure adopted was to average 
pixels on the DIRBE weekly maps in a circle of radius 1.5 degrees centred on 
each 
galaxy, after interpolation to the epoch of each ISOPHOT measurement. The
average of these brightnesses was then used to construct a colour-corrected 
SED 
of the typical background emission encountered in the cluster in the  
25, 60, 100, 140 and 240\,${\mu}$m DIRBE bands (Fig.~5). 
%This SED was fitted 
%as the sum of a Planck component (for the Zodiacal light 
%with the dust temperature $T_{\rm D}=320$\,K), and of two modified 
%($m=2$) Planck 
 %components (with best fit paramenters $T_{\rm D}=30$ and 16\,K for the warm 
%and cold dust emission from the galaxy, respectively).

Next, the ISOPHOT measurements of the background towards each galaxy 
were extracted as weighted averages of the P32 maps outside the region 
enclosed by the 25.5 mag. blue isophote, and colour corrected according 
to the fitted SED of Fig.~5. The colour corrected ISO backgrounds are plotted
versus the colour corrected DIRBE measurements towards each galaxy 
for each ISOPHOT filter in Fig.~6. Though a good correlation is seen
between ISO and COBE-DIRBE at 60\,${\mu}$m, there is a scatter in the
ratios, the dispersion becoming more pronounced with increasing wavelength. 
This might be due either to variations in detector responsivity between 
different ISOPHOT observations or due to the effect of structured Cirrus 
emission. We believe that
the latter is the most likely explanation, as responsivity variations would
be expected to affect both the C60 and C100 measurements in equal measure.
Furthermore, particularly in the C160 band, measurements with the highest 
ISOPHOT/DIRBE ratios are those with the largest dispersion in the background
signal between different detector pixels, which one would expect when viewing
regions of brighter than average structured Cirrus. Thus, the length of the
1$\,\sigma$ bars in the C160 ISO versus DIRBE comparison in Fig.~6c may give 
an indication of the level of Cirrus confusion in the C160 band.

The systematic differences in the response of ISOPHOT and COBE-DIRBE to the
backgrounds are summarised in Table~3.
The third column of the table gives brightness values for the fit (Fig.~5) 
to the mean background, interpolated to the
reference wavelengths of the ISOPHOT C60, C100 and C160 filters. The
fourth column gives the mean of the ratios of the colour corrected responses
of the two instruments to the backgrounds.

%%%%%%%%%%%%%%%%%%%%%%%%%%%  Table 3  %%%%%%%%%%%%%%%%%%%%%%%%%%%% 
\begin{table} 
\tablenum{3}
\caption{Comparison of ISO and DIRBE response to the background}
\begin{tabular}{lccccccc}
\hline
Filter  &   $\lambda\_{\rm ref}$ &  Brightness & ISO/DIRBE \\ 
  & (${\mu}$m) & (MJy/sr) &  & \\
\hline 
C60 & 60 & 14.59 & 0.984 \\
C100 & 100  & 8.54 & 0.844 \\
C160 & 170 & 6.07 & 1.130 \\
\hline
\end{tabular}
\end{table}

%%%%%%%%%%%%%%%%%%%%%%%%%%%%%%%%%%%%%%%%%%%%%%%%%%%%%%%%%%%%%%%%%%%%%%%%%%%%%%

Prior to further analysis all ISO data was multiplied by factors of 
1.016, 1.185 and 0.885 in the C60, C100 and C160 bands, respectively, 
in order to bring the ISO measurements onto the COBE flux scale. 
This was done primarily because of the remaining intrinsic
uncertainties in the absolute ISO calibration.
In particular uncertainties in the ramp non-linearity correction for short reset 
intervals make this approach almost unavoidable. The adoption of the COBE 
flux scale also 
provides a basis for cross-calibrating the C100 and C160 detectors, since 
the spectral coverage of the DIRBE instrument encompasses that of ISOPHOT.
The adoption of the COBE-DIRBE flux scale raises 
appreciably the 100/170 colour temperature of the galaxies in the survey
compared with the pure ISO calibration.
 
This procedure based on 
the respective responses of ISO and COBE-DIRBE to a smooth background does 
not necessarily imply a correct adoption of the COBE-DIRBE flux scale for 
discrete sources. In common with the flux scales of IRAS and COBE-DIRBE, the 
ISO flux scale is fundamentally based on staring observations of point-like
sources (stars, asteroids and planets in the case of ISOPHOT-C), which are
subsequently converted into a calibration of surface brightness using
measured values for the beam solid angles.
To check the validity of the conversion of the ISO flux scale to the 
COBE-DIRBE flux scale, as described above, a 
comparison was made between the response of ISO and IRAS to 
bright Virgo cluster galaxies in their respective 60 and 100${\mu}$m
band-passes.

\subsection{Comparison of ISO and IRAS photometry of bright galaxies}

33 of the 63 galaxies in the ISO sample were detected by IRAS.
Although most of these sources were resolved by ISO, 
they are sufficiently compact (typical IR extent of order of the FWHM of the 
ISO point spread function) to provide an objective check of the response of
the C100 detector to discrete sources. Spatially integrated fluxes were 
extracted from the ISO maps using the fitting procedure 
described in Sect.~6 and plotted against the IRAS values at 60 and 
100${\mu}$m from Table~1 in Fig.~7a,b. 

Except for a few bright galaxies, a good linear correlation is seen between 
the ISO and IRAS fluxes, consistent with values from the ratio of fluxes 
measured by ISO to those measured by IRAS of ${\rm ISO/IRAS}=0.95$ and 0.82 at 
60 and 100\,${\mu}$m respectively. 
The bright galaxies with 
ISO/IRAS ratios well below the correlation were extended objects not well 
covered in the cross-scan direction. In these cases we attribute the 
shortfall in integrated flux measured by ISO to the presence of extended 
emission outside the perimeter of the ISO maps, which could not be fully 
recovered in the integration of the Gaussian model fits (Sect.~6).

The mean ratio ISO/IRAS for the galaxies in our sample corresponds to the
response of ISO to discrete sources relative to that of IRAS, where the ISO
measurements have been scaled onto the COBE-DIRBE flux scale.
%The mean ratio ISO/IRAS for the galaxies corresponds to the response of
%ISO to discrete sources relative to that of IRAS after the scaling of the
%ISO measurements onto the COBE-DIRBE flux scale. 
These ratios for
discrete sources can be compared to the ratios
COBE-DIRBE/IRAS for the background, derived using
table IV.D.1 of Wheelock et al. (1994).
These background ratios are COBE-DIRBE/IRAS$\,=\,$0.87$\,\pm\,$0.05 and
0.72$\,\pm\,$0.07 in the 60 and 100\,${\mu}$m bands, respectively.
Thus, the ratios for the
background COBE-DIRBE/IRAS are the same (within the statistical uncertainties)
as the ratios ISO/IRAS for discrete sources (0.95 and 0.82, after scaling the
ISO measurements onto the COBE-DIRBE flux scale).
%Thus, within the statistical uncertainties the ratios for the
%background COBE-DIRBE/IRAS are the same as the ratios
%ISO/IRAS$\,=\,$0.95 and 0.82 for discrete sources after scaling the
%ISO measurements onto the COBE-DIRBE flux scale. 
This validates the
basic assumption used to convert the ISO measurements to the COBE-DIRBE
flux scale, namely that the scaling factors COBE-DIRBE/ISO
derived from background measurements are also valid for discrete sources.

Finally, we note that Odenwald, Newmark \& Smoot (1998) found the 
colour corrected ratio COBE-DIRBE/IRAS for bright galaxies measured by 
COBE-DIRBE to be $0.94\pm0.12$ and $1.13\pm0.16$ in the 60 and 
100\,${\mu}$m bands, respectively. A good agreement with 
the ISO/IRAS ratios for Virgo galaxies presented here is found in the 
60\,${\mu}$m band, but there is a marginal discrepancy in the 100\,${\mu}$m 
band,
where the Odenwald et al. results would suggest that the COBE-DIRBE/IRAS 
response is significantly greater for discrete sources than for  
the smooth background. We have no explanation for this discrepancy. However, 
we do not make explicit use of the COBE/DIRBE response to discrete sources
in converting the ISO data to the COBE-DIRBE flux scale.

\section{Extraction of photometry}
There are two methods to extract the photometry from the data, namely by
integrating the raw background-subtracted maps or by fitting a model to the
data and then integrating the model to infinity. We applied both methods to our
data and show that for this kind of observation the integration of the model 
to infinity is a more accurate method. 

\subsection{The integration of the raw maps}

The raw integration of the background-subtracted maps has some major 
disadvantages which make this
method less desirable for such analysis. 
%First of all this method will
%overestimate the flux densities for bright point sources due to the 
%undersampling, mainly in the cross scan direction, where the map pixel size 
%is $46^{\prime\prime}$
%at 60 and 100 ${\mu}$m and 92$^{\prime\prime}$ at 170 ${\mu}$m. 
First of all the
raw integration of the map will be affected by noise, especially for low
surface brightness objects. Secondly this method will not take into
consideration any emission that was not included in the scan. This would
affect mainly large objects and especially galaxies with large disks scanned 
along the minor axis, where the map covered only the
central part of the galaxy. The FIR emission beyond the scan is usually
recovered when fitting a model to the data, except for some extreme cases where
the undersampling will also affect the reliability of finding the right 
parameters for the fit. We nevertheless give flux densities obtained from 
integration of the raw maps as a first approximation measurement (Table~7).

\subsection{The integration of the model fit}

The model fit to the data is in principle a more accurate method for 
extracting the photometry, providing the model is a good description of the
true brightness distribution. This method also solves the problems related 
to the integration of the raw maps. 
In all cases we fitted 2-dimensional models to the maps, and all
models were Gaussians. However it was necessary to fit different combinations 
of models, according to the different morphologies encountered in the sample, 
such
as single Gaussian fit, two Gaussian fit, multi-component fit, and also a beam
fit for point-like sources. Based on the statistics given by the
error analysis, we decided on a case by case basis which model better fits 
the data. We
adopted the fitting routines from MINUIT (CERN Program Library) which is a tool
to find the minimum value of a multi-parameter function and analyze the shape
of the function around the minimum. This tool is especially suited to handle
difficult problems, like those which can be encountered in our 2-dimensional
maps. It gives very robust results in computing the best-fit parameter
values and uncertainties for non-linear problems, taking into account 
correlations between parameters. This is important in order to 
have a reliable estimate of the errors in the fitting parameters, to put 
upper limits on the non-detection, and to detect the presence of a disk or
nuclear component in the fit. The flux densities derived from integrating the 
model fit to infinity are listed in Table~7.

\subsubsection{Beam fit}

The first step in deriving the photometry is to find a good model for the
beam and to evaluate any departure between the model and the true beam.
In an ideal case one would have an analytical description of the true beam
shape and the integration under a fit to the observations of a true point
source would give exactly the same flux density as of the real
observation. Unfortunately there is no analytical description of the true
shape of the beam. Therefore we describe the beam on the drift-corrected maps 
as a 2-dimensional circular Gaussian. Because the Gaussian model differs in 
shape from the true beam profile, the integration to infinity of the best 
fit to maps is in general systematically different from the true flux 
densities. To account for this difference it is necessary to correct the 
model integrated fluxes by some correction factor, $f_{\rm corr}$. We refer to 
this correction as the ``beam 
correction''. To determine the value of $f_{\rm corr}$ we fitted simulated
noiseless maps of 1 Jy point sources, sampled in exactly the same way as 
the ISO Virgo maps, with circular Gaussians.
The widths of these gaussian beam fit models were 
${\sigma}=18.56^{\prime\prime}$,  ${\sigma}=19.95^{\prime\prime}$ and 
${\sigma}= 39.49^{\prime\prime}$ at 60, 100 and 170 ${\mu}$m, respectively,
where the Full Width at Half Maximum is ${\rm FWHM}=2.355 \times {\sigma}$. In
all cases the center of the Gaussian was fixed to the center of the map,
such that the only free parameter of the fit was the amplitude of the
Gaussian. An example of the model fit to the simulated beams is
given in Fig.~8.
The flux densities obtained from integrating the model fit to infinity 
were 0.87, 0.84 and 0.87\,Jy at 60, 100 and 170\,${\mu}$m, 
respectively.  The flux densities are always below 1\,Jy at all
three wavelengths, due to the differences between the assumed Gaussian 
model and the simulated beam. Thus a correction factor was assumed for all 
flux densities derived from model fits on point sources or on sources which 
include a point-like source component. 
The correction factors were $f_{\rm corr}=1.149$ at 60 and 170\,${\mu}$m and 
$f_{\rm corr}=1.19$ at 100 ${\mu}$m, where the uncorrected flux densities 
derived from integrating the model fit have to be multiplied with $f_{\rm corr}$.

When fitting real Virgo maps of point-like sources with 2-dimensional 
circular Gaussians we obtained an optimum width for the beam fit model of 
${\sigma}=17.41^{\prime\prime}$ at both 60 and 100\,${\mu}$m. The differences
between these fit parameters and those obtained for simulated beams are due to
the distortions induced in the beam profile by the responsivity drift
correction procedure (see Sect.~4.2).  For a few cases we found the need to 
model the beam with a Gaussian having even a smaller width, 
$\sigma=15.41\,^{\prime\prime}$ at 60 and 100\,${\mu}$m. These latter cases are
indicated in all tables with the comment ``modified beam''.
The effect of decreasing the width of the beam is hardly apparent for the
relatively fast C200 detector; therefore
at 170\,${\mu}$m we adopted the same width as for the simulated beams.  
Throughout this paper we refer to the 2-dimensional circular Gaussian model
with ${\sigma}=17.41^{\prime\prime}$ at 60 and 100\,${\mu}$m, and 
${\sigma}= 39.49^{\prime\prime}$ at 170\,${\mu}$m as the ``beam fit''.

The flux densities derived from integrating the beam fit model to infinity 
were then multiplied with the beam correction factors derived from the 
simulated beams. Presumably these correction factors will give only lower 
limits for the integrated fluxes, due to the change in the width of the 
derived profiles at 60 and 100\,${\mu}$m. However we expect the difference 
between the correction factors derived from simulated beams and the true 
correction factors which would need to be applied on real observations to be  
smaller than other uncertainties in the photometry, and we make no attempt to
correct for this effect.

The beam fit was utilised for a few galaxies which appeared
as point sources, and was also utilised as a way of deriving
upper limits for all our non-detections. All such cases
are commented as either ``beam fit'' or
``beam fit; non-detection'' in the listing of fit parameters
in Table~4.
%The beam fit was considered for a few cases of our Virgo galaxies which
%appeared as point sources and was also considered for all our non-detections,
%as a way of deriving upper limits. The parameters of the beam fit are 
%listed in
%Table~4, and those cases are commented as ``beam fit''. 
For all the other 
cases the data were fitted by a model which was convolved with the beam fit 
model. Model fits to the observed brightness profiles of all the Virgo galaxies
from our sample are presented in Fig.~10 \footnote{This figure is only
available for the journal paper.}. The brightness scale has been 
transformed to the COBE-DIRBE scale as described in Sect.~5. 
%Statistical 
%uncertainties for selected map pixels near the extremities and centre of the 
%scans are indicated by the $3\sigma$ error bars. The legend is as given in 
%Fig.~8.  

\subsubsection{One Gaussian fit}

Most of our data were fitted by a 2-dimensional elliptical Gaussian, obtained
by convolving a model source brightness distribution for each galaxy with a
beam fit model, as described in the previous section. The brightness
distribution of the source was itself modeled as a 2-dimensional elliptical
Gaussian, constrained to have the position
angle of the major axis as given by the B band isophotal photometry from
the VCC catalogue. Another constraint was to fix the major to minor axial
ratio again from the optical data (the VCC catalogue). These constraints were
needed because for most of the observations the map extent was only 2 or 3
pixels in the cross-scan direction, and also because of the  undersampling 
problem, which made impossible a reliable determination of the orientation 
and width of the Gaussian in the cross-scan direction. Such a constraint would 
nevertheless affect the photometry in cases where the FIR shape of the 
galaxies would not follow the optical shape of the galaxy.
 
The center of the Gaussian was fixed to the actual optical centre of the
galaxy. This was sometimes slightly offset from the geometrical center of the
map due to pointing offsets of the telescope. In many cases the optical
position within the map proved to be a poor representation of the actual center
of the galaxy in the FIR maps. Therefore an initial fitting routine was used 
to find the best position of
the Gaussians, and this refined position was finally adopted for the
fit. Larger shifts from the optical position were mainly found in the 
C160 maps, probably due to the undersampling of the maps coupled with
variation in the responsivity within one map pixel. 
It is also not unreasonable to consider that part of the shifts are due to 
real offsets of the optical from the FIR center of the galaxies.  
The fitted parameters are the amplitude and the width ($\sigma$) of the 
Gaussian. The parameters of the fit are listed in Table~4. 

\subsubsection{Two Gaussian fit}

In several cases there was evidence for a nuclear component plus a disk 
component,
which we fitted using a superposition between a beam fit model and a 
2-dimensional 
elliptical Gaussian convolved with the beam fit model. In this case the PA 
and axial ratio of the disk component were
again fixed from the optical data, and the fitted parameters were the amplitude
of the nuclear component (the beam fit model) and the amplitude and width of 
the elliptical Gaussian representing the disk. Again the parameters of the 
fit are listed in Table~4.  

\subsubsection{Multi-component fit}

In a few cases some morphological features were observed in the maps,
and therefore a more sophisticated fit was required. Thus for VCC~66, the 
maps
at 60 and 100 microns revealed the presence of a peripheral  HII region 
complex, which we had to
fit with a third Gaussian. The galaxies VCC~1554 and VCC~1727 had also to be 
fitted with a third Gaussian, this time to account for extended emission.
The galaxies with multi-component fits are marked in Table~4  as such, and the
actual parameters of the fit are given in Table~5. 

A special case is the interacting system VCC~1673/VCC~1676. 
At 60 and 100\,${\mu}$m
the observations were twice as finely sampled in
the cross-scan direction, resulting in a
higher resolution map of the system. These maps could not be 
fitted by a model, due to the detailed morphologies, and are better  suited 
to the direct integration of the raw maps. Therefore this interacting pair is 
not included in Fig.~10, but rather will be the object of a separate paper. 
At 170\,${\mu}$m the sky was undersampled in the cross-scan direction in the
same way as for the rest of the galaxies, and thus a model including Four 
Gaussians was considered. The parameters of this fit are given separately in 
Table~6. 

\subsubsection{The ${\chi}^2$ statistics}

The ${\chi}^2$ statistics reflects the random errors in the fit, being an 
indicator
of the goodness of the fit. In our case ${\chi}^2$ is also affected by the
undersampling in the cross-scan direction. This effect appears simply because the
pixel map is supposed to be uniformly illuminated, while our Gaussian model
assumes an illumination variation within the pixel. This will be interpreted as
a departure between the
data and the fit, resulting in an increased ${\chi}^2$. For low flux 
densities the ${\chi}^2$ will be dominated by the random errors, and the 
effect of the undersampling is negligible. But with increasing source 
brightness ${\chi}^2$ will start to be dominated by the errors resulting 
from the undersampling, and thus artificially large numbers of the ${\chi}^2$ 
will be obtained. This effect is illustrated in Fig.~9 for the flux densities 
at 60 micron. As expected, for bright sources there is a good linear 
correlation between the flux densities derived from integrating the model to 
infinity and 
the ${\chi}^2$ of the fit. For faint sources we expect the ${\chi}^2$ to be 
dominated by random errors, and thus a scatter diagram is obtained on the 
left hand side of the plot. For bright sources the real departure between the 
Gaussian model and the data is given by the scatter in the linear correlation.
 Thus, for bright sources, the values of the
${\chi}^2$ had to be corrected for the undersampling effect. The following
correlations were used for the correction:

\begin{eqnarray}
{\chi}^2  =  {\rm F}_{\rm model}\times 0.70-0.5;  & {\rm at}\,\,  60\,{\mu}{\rm m}\\
{\chi}^2  =  {\rm F}_{\rm model}\times 0.40-0.2;  & {\rm at}\,\, 100\,{\mu}{\rm m}\\
{\chi}^2  =  {\rm F}_{\rm model}\times 0.25-0.2;  & {\rm at}\,\, 170\,{\mu}{\rm m}
\end{eqnarray}

The values of the ${\chi}^2$ corrected for the undersampling effect are listed
in Table~4 (and in Table~5 and 6 for multi-component fit).

\subsubsection{Upper limits}

In all cases where we had non-detections we derived upper-limits based on beam
fits to the data, on the assumption that the galaxy which escaped detection was
a point-like source. The upper limits are thus the flux densities 
corresponding to the 3 $\sigma$ uncertainties in the amplitude of the beam 
fit model, where the center of the gaussian model was fixed, as described in 
the previous section. These cases are marked in Table~4 as \lq\lq beam fit; 
non-detection\rq\rq.

We note that the assumption that the undetected
galaxies are  point-like sources may not always be true.
For low surface brightness extended galaxies the
upper limits obtained with our method could be lower
than the actual integrated flux density.

\section{Detection statistics and basic FIR properties of the sample}

From the observed 63 galaxies (61 galaxies observed at all three FIR 
wavelengths and 2 galaxies observed only at 100 and 170\,${\mu}$m) we have 
detected 54 (85.7$\%$) galaxies at least at one wavelength and 40 
(63.5$\%$) galaxies at all three wavelengths. 9 (14.3$\%$) galaxies were not 
detected at two wavelengths, namely at 60 and 100\,${\mu}$m, and 5 
(7.9$\%$) galaxies were not detected at one wavelength. The 9 galaxies not
detected at any wavelength are mainly early type 
spirals, of SB0 (VCC~984), SBa (VCC~1047) or Sa (VCC~1158) Hubble type, and Im 
irregulars (VCC~17, VCC~169, VCC~666, VCC~1121). There is also one SBm 
(VCC~1217)
and one Scd: (VCC~1569) which escaped detection. The non-detected early type
spirals are bright objects ($12.1<B_{\rm T}<12.8$) in the cluster core 
while three of the non-detected Im are faint galaxies ($15.2<B_{\rm T}<16.8$) 
in the cluster periphery.

The statistics of the detections at each wavelength are given in Table~8. 

\begin{table}[htb]
\tablenum{8}
\caption{Statistics of the detections at individual wavelengths}
\begin{tabular}{r|lll}
\hline
 & 60\,${\mu}$m & 100\,${\mu}$m & 170\,${\mu}$m \\
\hline
detected & 41 & 43 & 54\\
         & (67.2$\%$)     & (68.3$\%$)    & (85.7$\%$)\\
\hline
observed & 61              & 63              & 63             \\ 
\hline
\end{tabular}
\end{table}

The faintest B band galaxy from our sample which was detected at 
least at one FIR wavelength was VCC~1001. This is an Im galaxy in the cluster 
core with ${\rm B}_{\rm T}=16.6$ and was detected only at 170\,${\mu}$m. 
(The galaxy was not observed at 60\,${\mu}$m and was not detected at 
100\,${\mu}$m.) The derived flux at 170\,${\mu}$m is $0.23\pm0.03$\,Jy. The
galaxy has an extremely cold dust emission and we will discuss its properties 
in a subsequent paper. The faintest B band galaxies from our sample which 
were detected at
least at two FIR wavelengths were VCC~130 and VCC~1750, two BCDs in the cluster
periphery with ${\rm B}_{\rm T}=16.5$. Both galaxies were detected only at 
100 and 170\,${\mu}$m. VCC~1750 reached $0.04\pm0.01$ and $0.08\pm0.02$\,Jy at 
100 and 170\,${\mu}$m, respectively.

The faintest detected emission at 60\,${\mu}$m is $0.05\pm0.01$\,Jy 
(detected in 
VCC~1675, a peculiar galaxy in the cluster periphery, and in VCC~1725, a 
Sm/BCD galaxy also in the cluster periphery). 
%The next galaxy with very faint detected emission at 60\,${\mu}$m (0.06\,Jy), 
%is VCC1419, a  ``Spec dust'' classified galaxy within the cluster core. 
The faintest detected emission at 100\,${\mu}$m is $0.04\pm0.01$\,Jy 
(detected in VCC~1750, one of the two galaxies which were also assigned as 
being the faintest B band galaxies detected at least at two FIR wavelengths). 
%Very faint emission at 100\,${\mu}$m (0.05\,Jy) is also found for VCC~848, an 
%``Im\,pec/BCD'' galaxy in the cluster periphery. 
Finally the faintest detected emission at 170\,${\mu}$m  is $0.08\pm0.02$\,Jy 
(detected in VCC~130 and in VCC~1750, which were also the faintest B band 
galaxies detected at least at two FIR wavelengths). For comparison the 
faintest detected IRAS emission on the galaxies from our sample is 0.2\,Jy 
(VCC~459, VCC~1410, VCC~1757) at 60\,${\mu}$m and 0.4\,Jy (VCC~1757) at 
100\,${\mu}$m. Thus our survey can go $10-20$ times deeper (at 100\,${\mu}$m) 
than the IRAS survey. 

The faintest upper limits (defined as 3\,${\sigma}$ errors) assigned to the
galaxies in our sample are 0.03\,Jy at 60\,${\mu}$m (VCC~169, VCC~666, 
VCC~1581, VCC~1750), 0.02\,Jy at 100\,${\mu}$m (VCC~1001, VCC~1121) and 
0.04\,Jy at 170\,${\mu}$m (VCC~666). For comparison the faintest upper
limits assigned by IRAS to the galaxies from our sample are 0.2 and 
0.5\,Jy at 60 and 100\,${\mu}$m, respectively. Thus our survey 
assigns upper limits which are again $\sim20$ times fainter (at 100\,${\mu}$m) 
than those assigned by the IRAS survey. The averaged 3\,$\sigma$ upper 
limits of the galaxies (point sources) from our sample  were 43, 33 and 
58\,mJy at 60, 100 and 170\,${\mu}$m, respectively.

Interestingly, the highest detection rates were for the C160
band, even for dwarfs, despite the detection limit in terms of Jy being about
twice as bright as the C100 and C60 detection limits. Particularly for BCDs it
was not obvious that this was to be expected.
 
The galaxies with the brightest FIR fluxes are VCC~1673/VCC~1676, which form 
an interacting system with total integrated flux densities of $11.32\pm0.19$, 
$29.34\pm0.46$ and $96.85\pm1.56$\,Jy at 60, 100 and 170\,${\mu}$m, 
respectively.

From the observed 63 galaxies, 21 (33.3$\%$) galaxies 
present evidence for a nuclear and a disk component at least at one 
wavelength and 4 (6.3$\%$) galaxies present evidence for a multi-component 
morphology at least at one wavelength.

The flux density distributions at 60, 100 and 170\,${\mu}$m are given in
Fig.~11
. There is an obvious shift of the right hand side of the 
histogram towards larger fluxes with increasing wavelength. The distribution
at 170\,${\mu}$m is slightly broader than the distribution at 60, and 
100\,${\mu}$m due to the larger completeness of the sample at longer
wavelengths. Some faint galaxies detected at 170\,${\mu}$m failed detection at 
60\,${\mu}$m, producing the cut in the faintest side in the histogram at 
60\,${\mu}$m.

\section{Summary}

63 spiral, irregular and dwarf galaxies in the Virgo cluster
have been observed down to the limiting sensitivity
of the ISOPHOT instrument on board ISO in band-passes centred
on 60, 100 and 170 micron. Rapid oversampled scans covering
the entire optical extent of each target down to the
25.5\,mag\,arcsec$^{-2}$ B-band isophote and adjacent
background directions were made using ISOPHOT's focal
plane chopper in conjunction with a spacecraft raster.
For the first time, data taken in this mode
could be corrected for the complex non-linear response to the illumination
history in each observation. This allowed robust integrated photometry
and structural information to be extracted down to the confusion
limit in the 170 micron band and to the sensitivity limits
imposed by low-level glitching in the 60 and 100 micron bands.

The photometry in the 60 and 100\,${\mu}$m bands 
was found to be well correlated with the corresponding IRAS measurements 
for the 33 of the 63 galaxies detected by IRAS, with relative 
gains ISO/IRAS\,=\,0.95 and 0.82 at 60 and 100\,${\mu}$m respectively.

The faintest detected emissions from our galaxy sample were 50, 40 and 
80\,mJy at 60, 100 and 170\,${\mu}$m, respectively. The faintest 3\,$\sigma$ 
upper limits for integrated flux densities were 30, 20 and 40\,mJy 
(at 60, 100 and 170\,${\mu}$m). The averaged 3\,$\sigma$ upper limits (of 
galaxies with pointlike source appearance) were 43, 33 and 58\,mJy (at 60, 
100 and 170\,${\mu}$m). 54 galaxies (85.7$\%$)
were detected at least at one wavelength, and 40 galaxies 
(63.5$\%$) were detected at all three wavelengths. The highest
detection rate (85.7$\%$) was in the 170\,${\mu}$m band. 

From the observed 63 galaxies, 21 (33.3$\%$) 
presented evidence for a nuclear and a disk component at least at one 
wavelength and 4 (6.3$\%$) galaxies presented evidence for a multi-component 
morphology at least at one wavelength.

\vspace{1cm}

The data presented in this paper were taken in guaranteed observing
time made available by the ISOPHOT PI D. Lemke and the ISO Project Scientist 
M.F. Kessler. We thank U. Klein, J.Lequeux and B. Binggeli for advice and 
discussions concerning the selection of galaxies observed, D. Skaley for 
assistance in entering the observing parameters into the uplink database,
S. Niklas for support in the initial stages of the data evaluation,
P. Abraham and the ISOPHOT data centre for advice and assistance in the 
photometric comparison between ISO and COBE/DIRBE, and 
the staff of the ISO data centre at Villafranca for
advice on calibration and data analysis issues. RJT thanks
B. Madore for hospitality at the Observatories of the Carnegie 
Institution of Washington during the preparation of this paper. This project 
was supported by grant 50 QI 9201 of the DLR.

This research has made use of the NASA/IPAC Extragalactic
Database (NED) which is operated by the Jet Propulsion Laboratory,
California Institute of Technology, under contract
with the National Aeronautics and Space Administration.

\begin{figure}[htp]
\includegraphics[scale=1.0]{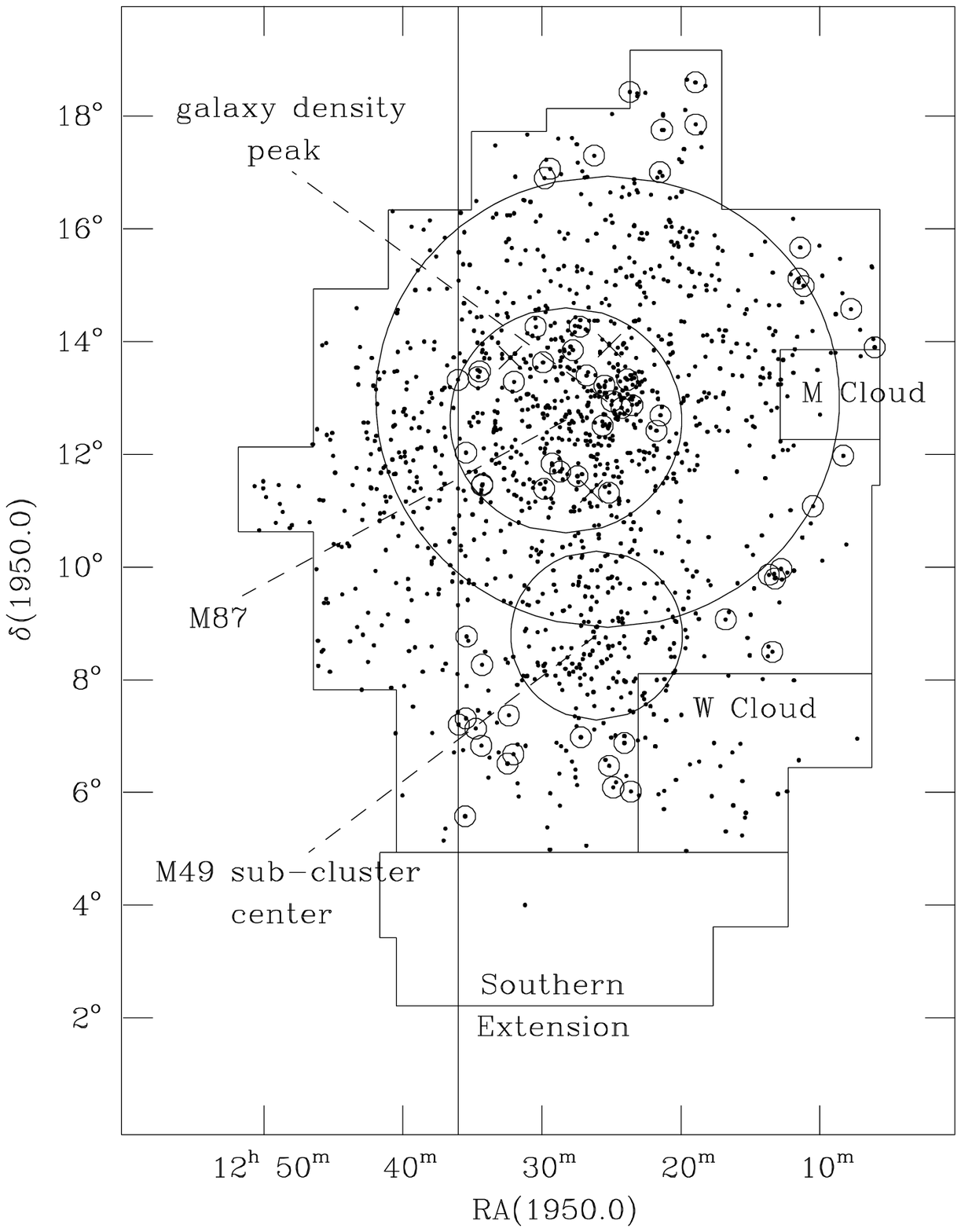}
\caption{Plot of the VCC member galaxies (dots), as defined by
BST85 and Binggeli, Popescu \& Tammann (1993),
with the cluster topology as given by Sandage, Binggeli \& Tammann (1985).
The 60 VCC galaxies of the statistical sample and the 3 VCC galaxies
of the extended sample observed by us are marked with open circles
and crosses, respectively. The Figure also show
the $\rm 4^o$ radius region centered on the maximum projected galaxy density
of the cluster (Binggeli, Sandage \& Tammann 1987), the $\rm 2^o$ radius 
region centered on M\,87, the $\rm 1.5^o$ radius region centered on 
the position of the maximum projected density of the M\,49 sub-cluster and 
the boundary line at $\rm RA(1950.0)=12.596^h$ (see text).}

\end{figure}

\begin{figure}[htp]
\includegraphics[scale=0.6]{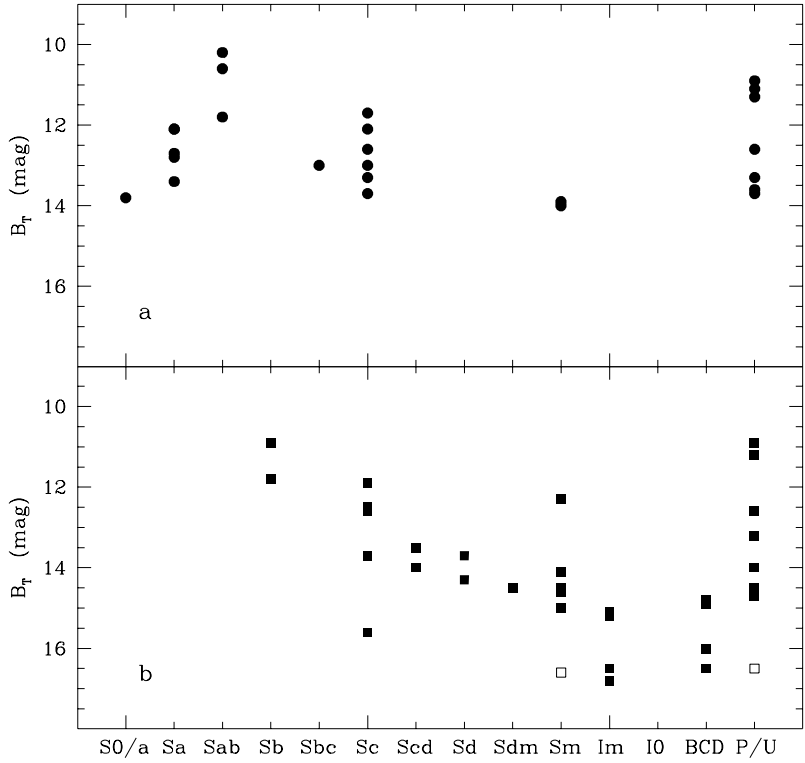}
\caption{The galaxy bivariate distribution in the plane
defined by morphological type and total apparent magnitude $\rm B_T$,
as listed in the VCC, for the core subsample ({\bf a})
and the periphery subsample ({\bf b}).
In each panel, the last morphological bin ``P/U'' contains either peculiar 
galaxies or objects of uncertain morphology. 
Circles and squares denote objects in the core and periphery
subsamples, respectively. Empty squares 
denote galaxies in the cluster periphery subsample not observed
by ISOPHOT (see text).}
%Empty circles and squares define objects in the core
%and in the periphery of the cluster, respectively,
%while filled symbols denote those observed with ISOPHOT.}
\end{figure}

%\begin{figure}[htp]
%\includegraphics[scale=1.0]{figure_3.eps}
%\caption{Differential galaxy distributions
%in total apparent magnitude $\rm B_T$ ({\bf a})
%and in Hubble type ({\bf b}) for the total VIRGO-ISO sample
% (black line) and the observed VIRGO-ISO sample (red line).}
%\end{figure}

%\begin{figure}[htp]
%\includegraphics[scale=1.0]{figure_4.eps}
%\caption{Logarithmic plot of $\rm f(60)_{IRAS}$ vs.
%$\rm f(100)_{IRAS}$ for the core sample ({\bf a})
%and the periphery sample ({\bf b}).
%Only the 44 objects detected by IRAS at both wavelengths are shown.
%In each panel, the solid line reproduces the mean IRAS far-IR colour
%$\rm f(60)/f(100)_{IRAS} = 0.4$ (with a dispersion of 0.12),
%found for the almost 2400 {\it normal} late--type galaxies
%of the UGC redshift sample (Bothun et al. 1989).
%The two short-dashed lines represent the $\pm$ 1-$\rm \sigma$
%values of $\rm f(60)/f(100)_{IRAS}$ listed in the latter study.
% }
%\end{figure}

\begin{figure}[htp]
\includegraphics[scale=0.6,angle=0.]{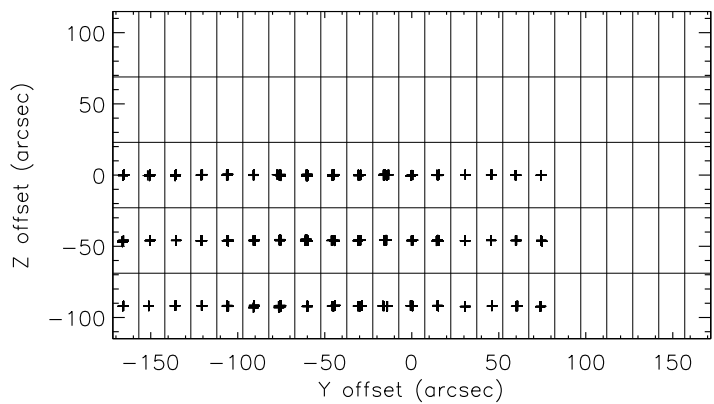}
\caption{Pointing directions observed towards VCC~1 at 60${\mu}m$ in detector 
pixel 1. The spacecraft raster dimension was 2x3 in YxZ.
Each pointing direction for 3509 individual data samples for the central
pixel of the C100 detector array is plotted as a cross. The rectangular grid
is the ``P32 natural grid'' (see text) for this observation, on which
the sky brightness distribution is to be solved. The grid sampling
is 14.94x45.95 arcsec. The registration of chopper and spacecraft pointing,
and the inherent pointing stability of ISO allow the data samples to be
typically within 1 arcsec of the centre of each pixel of the P32 natural
grid.
}
\end{figure} 

\begin{figure}[htp]
\includegraphics[scale=0.6]{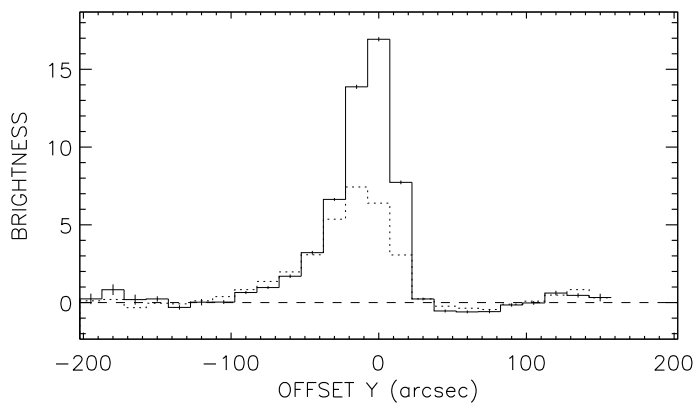}
\caption{Brightness profiles along the Y spacecraft scanning direction (raster
and chopper scan from right to left) through
the standard star HR1654 at 100${\mu}$m. The solid line represents data 
analysed in the same manner as the target Virgo cluster galaxies, while
the dotted line represents data with identical processing but without the 
responsivity drift correction.
}
\end{figure} 

\begin{figure}
%\plotone{backgroundfit.eps}
\includegraphics[scale=0.6]{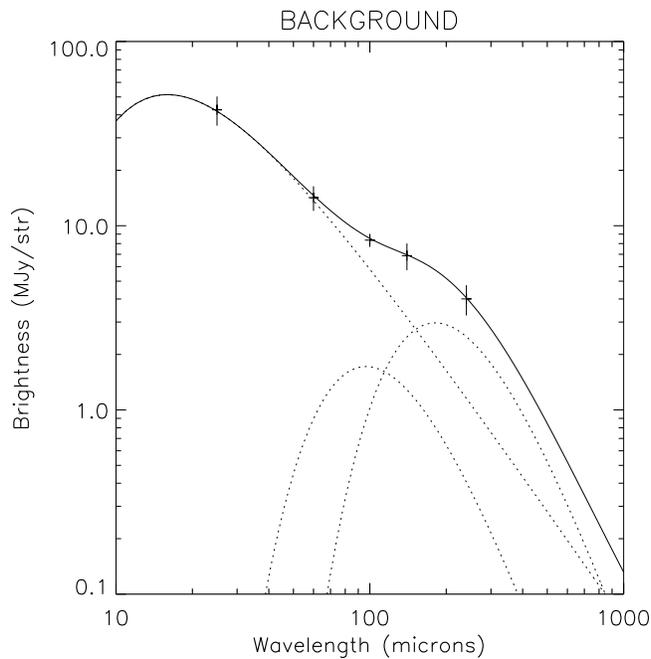}
\caption{The SED of the background 
emission seen towards the positions of the target galaxies, as measured 
by COBE-DIRBE. The vertical bars delineate the $1\,{\sigma}$ spread in 
DIRBE brightnesses measured at the target galaxies, and the horizontal
line denotes the average of these measurements. The solid line
represents a fit to the average brightnesses composed of the sum of 
three spectral components denoted by the three dotted curves: 
a black body spectrum with T=320K for the zodiacal component, 
and modified (m=2) black body spectra with 
T=30 and T=16 K for the foreground galactic dust emission. 
The DIRBE data were colour corrected according to the fitted SED.}
\end{figure}

\begin{figure}
\subfigure[]{
\includegraphics[scale=0.5]{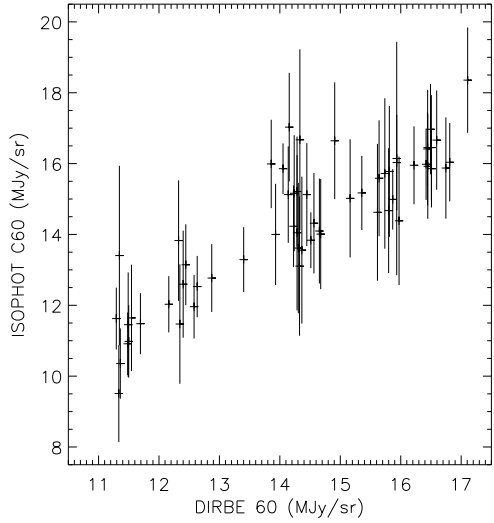}}
\subfigure[]{
\includegraphics[scale=0.5]{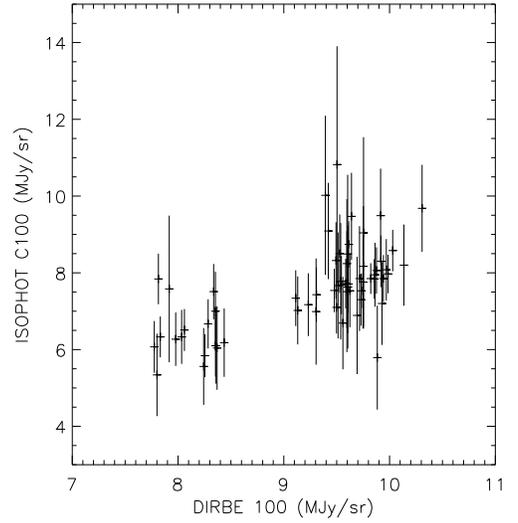}}
\subfigure[]{
\includegraphics[scale=0.5]{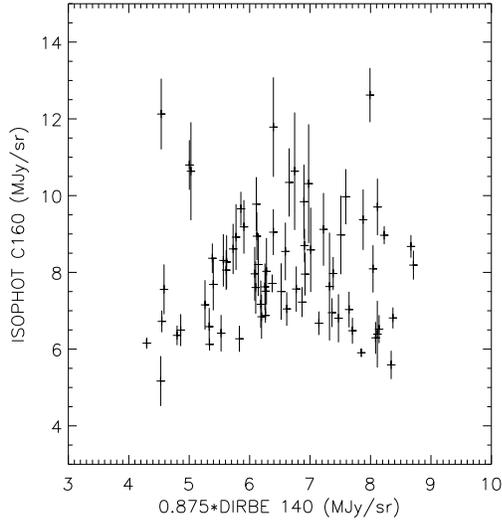}}
\caption{ISO backgrounds in the three ISOPHOT filters versus 
colour corrected brightnesses derived from DIRBE weekly maps in the
60, 100 and 140 ${\mu}m$ bands towards the 
positions of each target galaxy. The DIRBE measurements were
interpolated to the epoch of each ISO observations. Colour corrections 
were applied according to the 
fit to the mean DIRBE SED of Fig.~5. For the C160 filter, the DIRBE
data was multiplied by 0.875, which
is the factor by which the fit to the colour corrected background SED 
decreases between the DIRBE reference wavelength (140 ${\mu}m$) and the 
reference wavelength of the C160 filter (170 ${\mu}m$).
The vertical bars represent the $1\,{\sigma}$
 dispersion in the response
of individual detector pixels on ISOPHOT-C to the background.}
\end{figure}

%\begin{figure}
%\plotone{iso160_cobe140.eps}
%\includegraphics[scale=1.0]{iso160_cobe140.eps}
%\caption{}
%\end{figure}

\begin{figure}
%\plotone{c60_iras.eps}
\subfigure[]{
\includegraphics[scale=0.5]{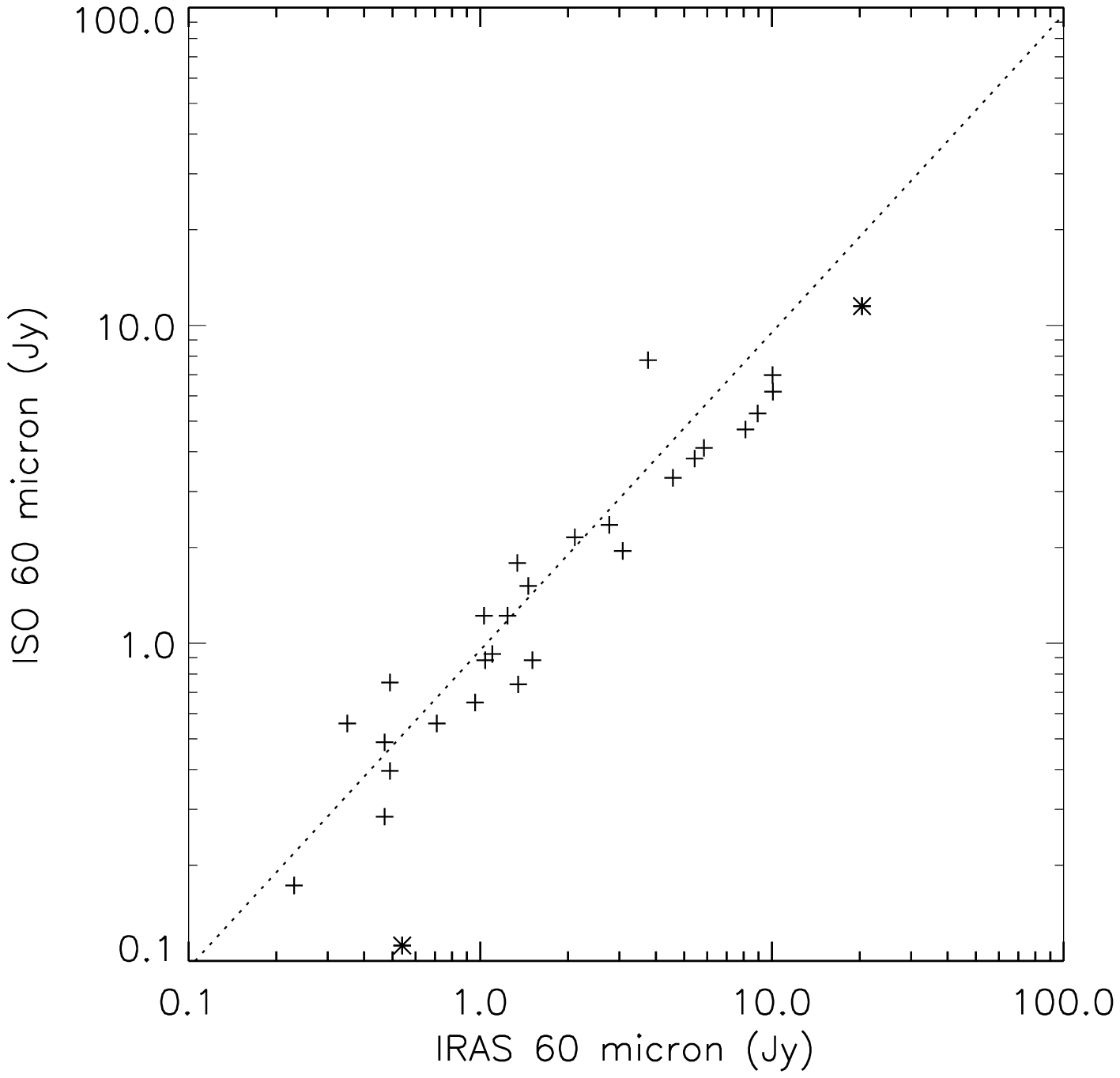}}
\subfigure[]{
\includegraphics[scale=0.5]{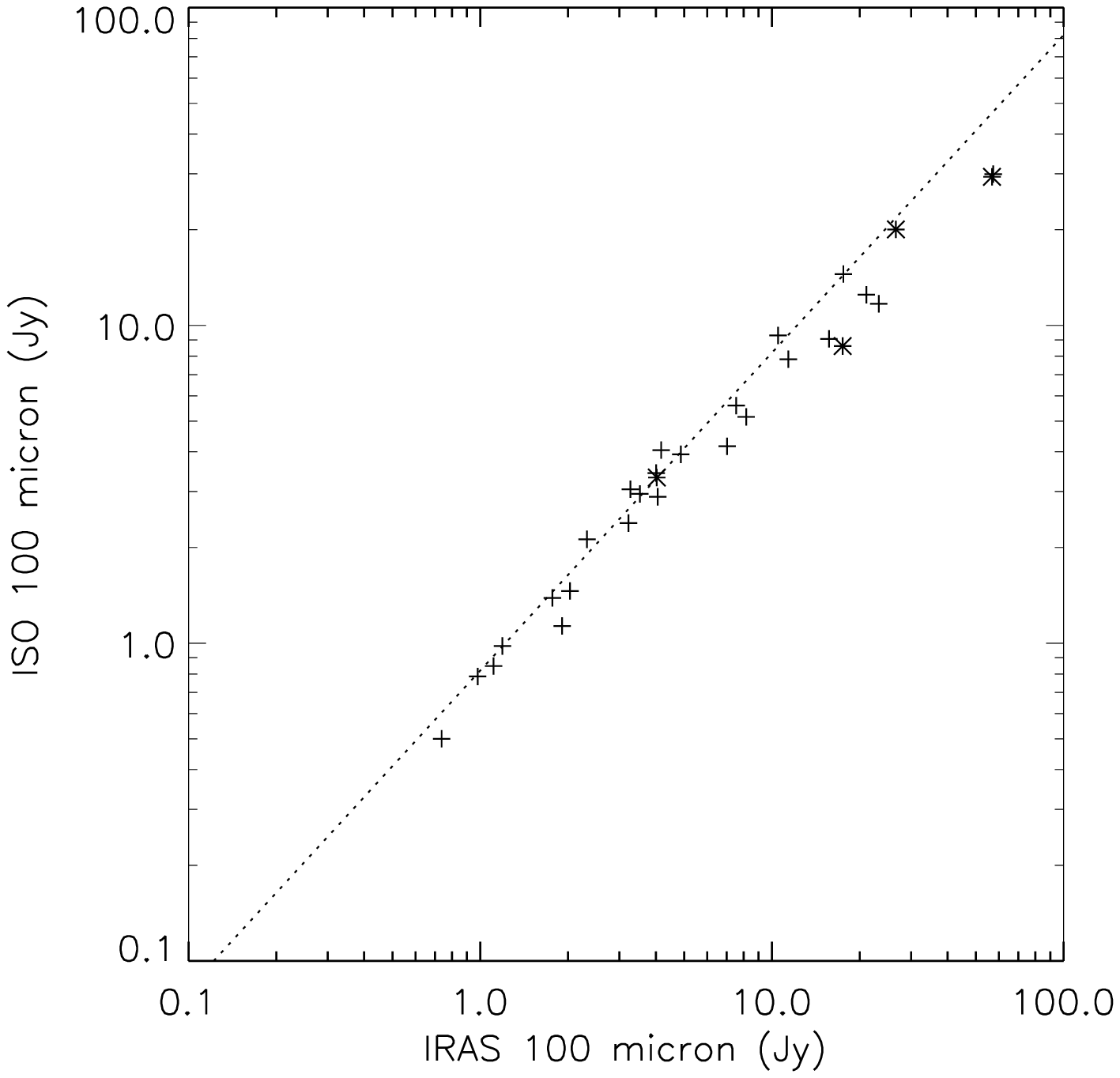}}
\caption{a) Integrated and colour corrected flux density of galaxies measured 
in the ISO C60 filter versus integrated colour corrected flux densities 
measured by IRAS in its 60\,${\mu}$m band. The colour correction 
factors were applied 
assuming the SED of the galaxies could be approximated by the superposition 
of two modified (m=2) Planck curves with temperatures 16 and 30\,K with the
same relative amplitude as used in Fig.~5 for the fitting of the 
foreground dust emission from the Milky way seen towards the Virgo cluster.
b) the same but for the 
ISO C100 filter and the IRAS 100\,${\mu}$m band. The ISO data represent the 
spatial integration of Gaussian components fitted as described in Sect.~6, and 
scaled to the COBE flux scale as described in Sect.~5. In each panel the 
dotted line is not a fit to the correlation, but represents the relation 
a) ISO/IRAS=0.95; b) ISO/IRAS=0.82 obtained in Sect.~5.2. The star 
symbols represent the 
integrated emission on the maps of a) VCC~1253, and VCC~1673; b) VCC~873, 
VCC~1379, VCC~1673 and VCC~1690, rather than the integration of the model 
fits. The galaxies plotted with stars have large $\chi$2 for the model fit (see
Table~4).}
\end{figure}

%\begin{figure}
%\plotone{c100_iras.eps}
%\includegraphics[scale=0.5]{c100_iras.eps}
%\caption{Integrated and colour corrected flux density of galaxies measured 
%in the ISO C100 filter versus integrated colour corrected flux densities 
%measured by IRAS in the 100 micron band. The ISO data represent the spatial 
%integration of gaussian components fitted as described in Sect.~6, and 
%have been scaled to the COBE flux scale as described in Sect.~5. The 
%dotted line is not a fit to the correlation, but represents the relation 
%ISO/IRAS=0.82. The star symbols represent the integrated emission on the 
%maps of VCC873, VCC1379, VCC1673 and VCC1690, rather than the integration 
%of the model fits.}
%\end{figure} 

\begin{figure}
\includegraphics[scale=0.85]{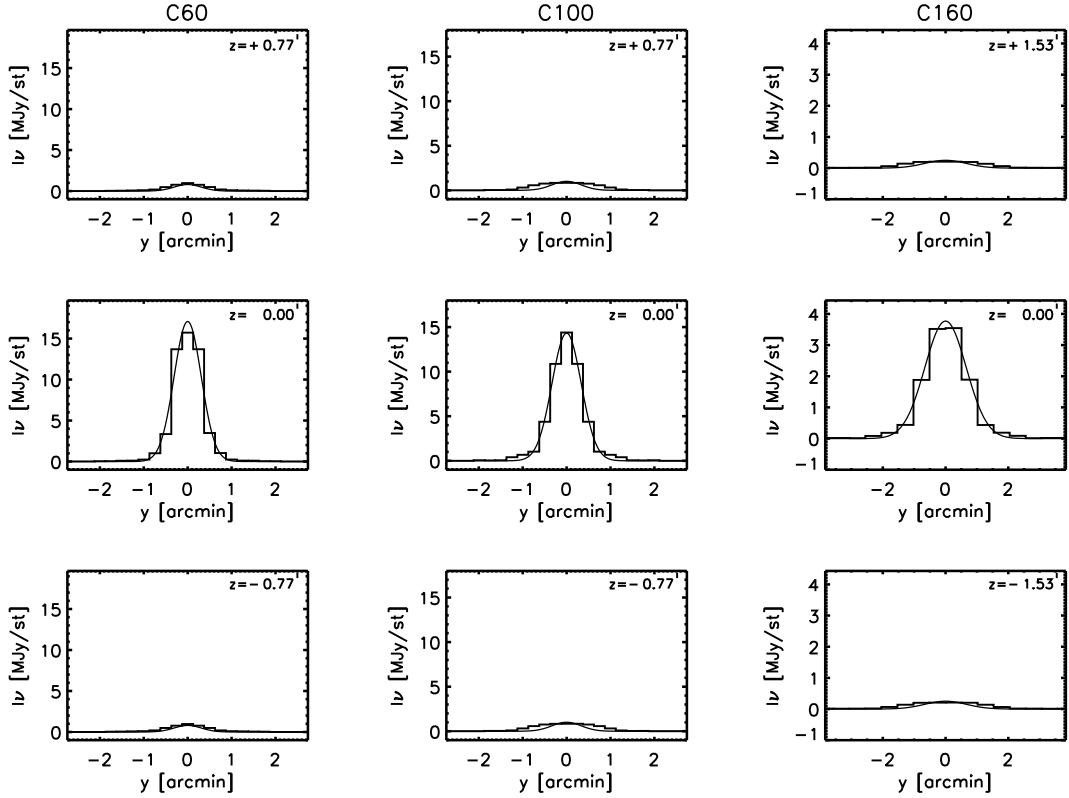}
\caption{Example of a beam fit on simulated noiseless maps of 1 Jy point 
sources, sampled in the exactly the same way as the ISO Virgo maps. The 
simulated beams are plotted as histograms
while the model fits are plotted as solid lines. The left, middle and 
right columns give the
fits to the 60, 100 and 170\,${\mu}$m maps, respectively. The middle row gives
the fits to the central map row (the profile along the scan direction), while 
the upper and the lower rows give the fits to the neighbouring map rows. The
shift in cross-scan direction z between the central and the neighbouring rows 
is indicated in each panel.}
\end{figure}

\begin{figure}
\includegraphics[scale=0.6]{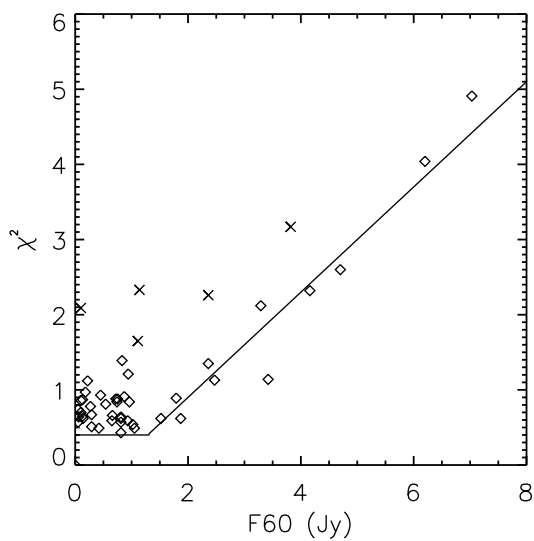}
\caption{The uncorrected ${\chi}^2$ values derived from fitting the 
Gaussian models to the maps
versus the corresponding flux densities at 60\,${\mu}$m obtained from 
integrating the
model fit to infinity. The data were plotted as diamonds, except for those
galaxies for which the model was an imperfect fit to the data, which were
plotted as stars. The solid line rising for ${\rm F}60>1.5$\,Jy represents the
linear correlations between
flux and ${\chi}^2$, as a result of systematic errors due to the
oversampling in cross-scan direction. For faint flux densities 
${\chi}^2$ is
dominated by systematic errors, producing a clumpy distribution at the left
hand side of the plot, with a lower envelope, which we traced by a solid line
to guide the eyes.}
\end{figure}

\clearpage

%\input{map1}
%\input{map2}

%\begin{landscape}
%\begin{figure}
%\hspace{-2.2cm}
%\includegraphics[scale=0.95, angle=270.]{f10_1.eps}
%\caption{Model fits to the observed brightness profiles of all the Virgo
%galaxies from our sample (except for the interacting system VCC1673/VCC1676 -
%see Sect.~6.2.4. Statistical uncertainties for selected map pixels near the
%extremities and center of the scans are indicated by the 3\,${\sigma}$ error
%bars. The legend is as in Fig.~8.}
%\end{figure} 
%\end{landscape}

%\begin{figure}
%\vspace{-0.8cm}
%\includegraphics[scale=0.95]{f10_2.eps}
%\end{figure} 

%\begin{figure}
%\vspace{-0.8cm}
%\includegraphics[scale=0.95]{f10_3.eps}
%\end{figure} 

%\begin{figure}
%\vspace{-0.8cm}
%\includegraphics[scale=0.95]{f10_4.eps}
%\end{figure} 

%\begin{figure}
%\vspace{-0.8cm}
%\includegraphics[scale=0.95]{f10_5.eps}
%\end{figure} 

%\begin{figure}
%\vspace{-0.8cm}
%\includegraphics[scale=0.95]{f10_6.eps}
%\end{figure} 

%\begin{figure}
%\vspace{-0.8cm}
%\includegraphics[scale=0.95]{f10_7.eps}
%\end{figure} 

%\begin{figure}
%\vspace{-0.8cm}
%\includegraphics[scale=0.95]{f10_8.eps}
%\end{figure} 

%\begin{figure}
%\vspace{-0.8cm}
%\includegraphics[scale=0.95]{f10_9.eps}
%\end{figure} 

%\begin{figure}
%\vspace{-0.8cm}
%\includegraphics[scale=0.95]{f10_10.eps}
%\end{figure} 

%\begin{figure}
%\vspace{-0.8cm}
%\includegraphics[scale=0.95]{f10_11.eps}
%\end{figure} 

%\begin{figure}
%\vspace{-0.8cm}
%\includegraphics[scale=0.95]{f10_12.eps}
%\end{figure} 

\clearpage

%\begin{figure}
%\vspace{-0.8cm}
%\includegraphics[scale=0.95]{f10_13.eps}
%\end{figure} 

%\begin{figure}
%\vspace{-0.8cm}
%\includegraphics[scale=0.95]{f10_14.eps}
%\end{figure} 

%\begin{figure}
%\vspace{-0.8cm}
%\includegraphics[scale=0.95]{f10_15.eps}
%\end{figure} 

%\begin{figure}
%\vspace{-0.8cm}
%\includegraphics[scale=0.95]{f10_16.eps}
%\end{figure} 

\begin{figure}
\subfigure[]{
\includegraphics[scale=0.5]{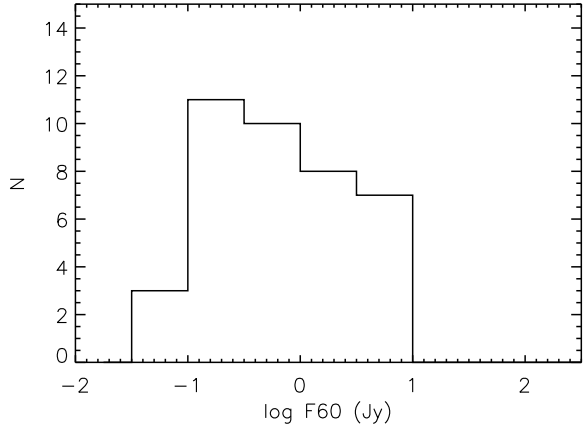}}
\subfigure[]{
\includegraphics[scale=0.5]{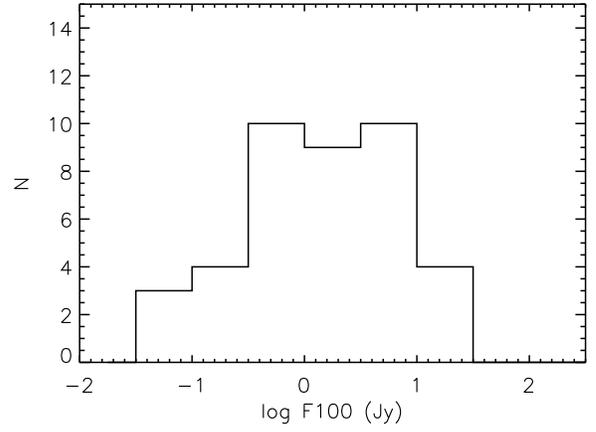}}
%\hspace{1cm}
\subfigure[]{
\includegraphics[scale=0.5]{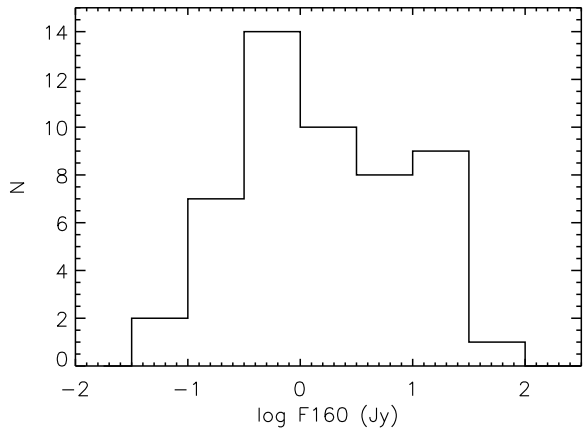}}
\caption{The flux density distributions at 60, 100 and 170\,${\mu}$m.}
\end{figure}

\input{tab1}
\input{tab2}
\input{tab4.tex}
\input{tab5_6.tex}
\input{tab7}

\clearpage

\end{document}

%% file: tab1.tex
%\makeatletter
%\def\jnl@aj{AJ}
%\ifx\revtex@jnl\jnl@aj\let\tablebreak=\\\fi
%\makeatother
%\pagestyle{empty}
%\ptlandscape
\begin{deluxetable}{llcrrrrrlrrrl} 
\tablecolumns{13}
\tableheadfrac{0.05}
\tabletypesize{\scriptsize}
\tablewidth{610pt}
\rotate
\tablenum{1}
\tablecaption{Galaxy parameters}  
\tablehead{\colhead{VCC} & \colhead{Other denomination} &
\colhead{C/P\tablenotemark{a}} & \colhead{$\alpha$(2000.0)} & \colhead{$\delta$(2000.0)} &
\colhead{D\tablenotemark{b}} & \colhead{R\tablenotemark{c}} & \colhead{PA\tablenotemark{d}} &
\colhead{type} & \colhead{$\rm B_T$} &
\colhead{$\rm F60$\tablenotemark{e}} & \colhead{$\rm F100$} &
\colhead{notes} \\
\colhead{} & \colhead{} & \colhead{} & \colhead{hh~mm~ss} &
\colhead{$\rm ^o$~$\rm ^{\prime}$~$\rm ^{\prime \prime}$} &
\colhead{$\rm ^{\prime}$} & \colhead{} & \colhead{deg} &
\colhead{} & \colhead{mag} & \colhead{Jy} & \colhead{Jy} &
\colhead{}}
\startdata
   1     &                                   & P & 12 08 20.3~~~& 13 41 03~~~~&  0.8~~& 4.3~~&   0~~& BCD? & 14.8~~& & & \\
  17     &                                   & P & 12 10 01.8~~~& 14 22 00~~~~&  1.4~~& 2.0~~&   0~~& Im & 15.2~~& & & \\    
  24     &                                   & P & 12 10 35.6~~~& 11 45 39~~~~&  0.8~~& 2.7~~&   0~~& BCD & 14.9~~& & & \\
  66     & NGC$\,$4178, UGC$\,$7215          & P & 12 12 46.5~~~& 10 52 06~~~~&  6.6~~& 2.8~~&  30~~& SBc & 11.9~~& 2.8~~~~& 9.0~~~~& HII \\
  81     & NGC$\,$4192A, UGC$\,$7223         & P & 12 13 26.2~~~& 14 46 20~~~~&  1.8~~& 1.2~~&   0~~& d:Sc & 15.6~~& & & \\ 
  87     &                                   & P & 12 13 41.3~~~& 15 27 13~~~~&  1.8~~& 2.0~~&   0~~& Sm & 15.0~~& & & \\
  92     & M$\,$98, NGC$\,$4192, UGC$\,$7231 & P & 12 13 48.3~~~& 14 54 01~~~~& 12.0~~& 3.7~~& 155~~& Sb: & 10.9~~& 6.5~~~~& 21.5~~~~& HII, LINER\tablenotemark{g,h} \\
 130     &                                   & P & 12 15 04.0~~~&  9 45 13~~~~&  0.4~~& 2.5~~&   0~~& BCD & 16.5~~& & & \\
\enddata
\end{deluxetable}

%% file: tab2.tex
%\makeatletter
%\def\jnl@aj{AJ}
%\ifx\revtex@jnl\jnl@aj\let\tablebreak=\\\fi
%\makeatother
%\pagestyle{empty}
%\ptlandscape
\begin{deluxetable}{rclrcccrl} 
%\tablecolumns{9}
\tableheadfrac{0.05}
\tabletypesize{\scriptsize}
\tablewidth{342pt}
\tablenum{2}
%\rotate
\tablecaption{Log-book of the observations}  
\tablehead{\colhead{VCC} & \colhead{TDT\tablenotemark{a}} &
\colhead{filter} & \colhead{raster\tablenotemark{b}} & \colhead{$\alpha$(2000.0)} &
\colhead{$\delta$(2000.0)} & \colhead{$\theta$\tablenotemark{c}} & \colhead{$\rm T_{int}$\tablenotemark{d}} &
\colhead{} \\
\colhead{} & \colhead{} & \colhead{} & \colhead{$\rm N_Y \times N_Z$} &
\colhead{hh~mm~ss} & \colhead{$\rm ^o$~$\rm ^{\prime}$~$\rm ^{\prime \prime}$} &
\colhead{deg} & \colhead{s}}
\startdata
1    & 22500605 & C60  & 2$\times$3~~~~ & 12 08 20.02 & 13 41 00.2 & 21.42 & 864~~~~ & \\
     & 22500605 & C100 & 2$\times$3~~~~ & 12 08 20.02 & 13 41 00.2 & 21.42 & 864~~~~ & \\
     & 22500606 & C160 & 3$\times$2~~~~ & 12 08 20.02 & 13 41 00.2 & 21.39 & 328~~~~ & \\
17   & 22500607 & C60  & 2$\times$3~~~~ & 12 10 01.87 & 14 21 42.5 & 21.36 & 864~~~~ & \\
     & 22500607 & C100 & 2$\times$3~~~~ & 12 10 01.87 & 14 21 42.5 & 21.45 & 864~~~~ & \\
     & 22500608 & C160 & 3$\times$2~~~~ & 12 10 01.87 & 14 21 42.5 & 21.35 & 328~~~~ & \\
24   & 23501521 & C60  & 2$\times$3~~~~ & 12 10 36.15 & 11 45 36.6 & 19.77 & 864~~~~ & \\
     & 23501521 & C100 & 2$\times$3~~~~ & 12 10 36.15 & 11 45 36.6 & 19.69 & 864~~~~ & \\
\enddata
\tablenotetext{a}{Target Dedicated Time (TDT) identifier;}
\tablenotetext{b}{raster pointing pattern (in spacecraft Y and Z);}
\tablenotetext{c}{position angle ($\theta$) of the spacecraft Y axis 
(positive East from North);}
\tablenotetext{d}{on-target integration time ($\ T_{\rm int}$);}
\tablenotetext{e}{${\dagger}$ indicates repeated observation.}
\end{deluxetable}

%% file: tab4.tex
\clearpage

\begin{deluxetable}{rrrrrrrrrrrl}
\tablenum{4}
\tabletypesize{\scriptsize}
\tablewidth{450.pt}
%\rotate
\tablecaption{The Fitting Parameters}
\tablehead{
\multicolumn{2}{c}{} & \multicolumn{3}{l}{One Gaussian Fit} &
\multicolumn{4}{l}{Two Gaussian Fit}\\
%\colhead{(1)} &
%\colhead{(2)} &
%\colhead{(3)} &
%\colhead{(4)} &
%\colhead{(5)} &
%\colhead{(6)} &
%\colhead{(7)} &
%\colhead{(8)} &
%\colhead{(9)} &
%\colhead{(10)} &
%\colhead{(11)} &
%\colhead{(12)} \\
\colhead{VCC}                       &
\colhead{$\lambda$}                 &
\colhead{S}                         &
\colhead{${\epsilon}$(S)}           &
\colhead{$\sigma$\tablenotemark{1}}                  &
\colhead{S$_{\rm n}$\tablenotemark{2}}               & 
\colhead{${\epsilon}$(S$_{\rm n}$)} &
\colhead{S$_{\rm d}$\tablenotemark{3}}               &
\colhead{${\epsilon}$(S$_{\rm d}$)} &
\colhead{$\sigma_{\rm d}$\tablenotemark{4}}          &
\colhead{${\chi}^2$}                &
\colhead{comments}                 \\
\colhead{}                          &
\colhead{${\mu}$m}                  &
\colhead{$\frac{\rm MJy}{\rm sr}$}  &
\colhead{$\frac{\rm MJy}{\rm sr}$}  &
\colhead{$^{\prime\prime}$}         &
\colhead{$\frac{\rm MJy}{\rm sr}$}  &
\colhead{$\frac{\rm MJy}{\rm sr}$}  &
\colhead{$\frac{\rm MJy}{\rm sr}$}  &
\colhead{$\frac{\rm MJy}{\rm sr}$}  &
\colhead{$^{\prime\prime}$}         &
\colhead{}                          \\
}
\startdata
   1&	60&   2.05&  0.33&   4.38&\nodata &\nodata&\nodata&\nodata&\nodata&  0.70&\\			    
    &  100&   2.39&  0.31&   9.87&\nodata &\nodata&\nodata&\nodata&\nodata&  0.52&\\
    &  170&   0.79&  0.11&  52.19&\nodata &\nodata&\nodata&\nodata&\nodata&  0.59&\\
  17&	60&   0.85&  0.29&\nodata&\nodata &\nodata&\nodata&\nodata&\nodata&  0.79& beam fit; non-detection\\
    &  100&   0.21&  0.27&\nodata&\nodata &\nodata&\nodata&\nodata&\nodata&  0.47& beam fit; non-detection\\
    &  170&   0.01&  0.10&\nodata&\nodata &\nodata&\nodata&\nodata&\nodata&  0.40& beam fit; non-detection\\
  24&	60&   0.68&  0.31&\nodata&\nodata &\nodata&\nodata&\nodata&\nodata&  0.59& beam fit; non-detection\\
    &  100&   0.79&  0.29&\nodata&\nodata &\nodata&\nodata&\nodata&\nodata&  0.25& beam fit; non-detection\\
\enddata

\tablenotetext{1}{The width ${\sigma}$ of the major axis of the elliptical
Gaussian source model in arcsec,
where the total width of the major axis of the fitted Gaussian is 
${\sigma}_{\rm total}^2={\sigma}^2+{\sigma}_{\rm PSF}^2$. The FWHM of the 
source model is $2.354\,{\sigma}$;}
\tablenotetext{2}{The amplitude S$_n$ of the nuclear component;
$^3$ The amplitude S$_d$ of the disk component;}
\tablenotetext{4}{The width ${\sigma}_{\rm d}$ of the disk major axis in
arcsec, where the total width of the major axis of the fitted Gaussian for 
the disk component is ${\sigma}_{\rm total}^2={\sigma}_{\rm d}^2+{\sigma}_{\rm PSF}^2$;}
\tablenotetext{5}{The parameters for the multi-component fit are listed in
    Tables~5,6.}
\tablenotetext{6}{VCC~159 shows evidence for faint extended emission at
160\,${\mu}$m, not fitted by the model. We only present a beam fit to the data,
resulting in a lower limit for the derived flux.}
\tablenotetext{7}{In some cases it was necessary to consider a narrower width
for the beam fit model, namely 
${\sigma}=15.41^{\prime\prime}$, ${\rm FWHM}=2.355 \times {\sigma}$ 
(see Sect. 6.2.1).}
\tablenotetext{8}{VCC~1196 presents evidence for extended asymmetric emission,
not fitted by the model. In this case the fluxes were derived only from the
integration of the raw maps.}
\end{deluxetable}

%% file: tab5_6.tex
\clearpage

\begin{deluxetable}{rrrrrrrrrrrl}
\tablenum{5}
\tabletypesize{\scriptsize}
\tablewidth{430.pt}
\rotate
\tablecaption{The Fitting Parameters for the multi-component fits}
\tablehead{
\multicolumn{2}{c}{} & \multicolumn{5}{l}{Main nuclear/disk component} &
\multicolumn{3}{l}{Other components}\\
%\colhead{(1)} &
%\colhead{(2)} &
%\colhead{(3)} &
%\colhead{(4)} &
%\colhead{(5)} &
%\colhead{(6)} &
%\colhead{(7)} &
%\colhead{(8)} &
%\colhead{(9)} &
%\colhead{(10)} &
%\colhead{(11)} &
%\colhead{(12)} \\
\colhead{VCC}                                  &
\colhead{$\lambda$}                            &
\colhead{S1$_{\rm n}$\tablenotemark{1}}                         &
\colhead{${\epsilon}$(S1$_{\rm n}$)}           &
\colhead{S1$_{\rm d}$\tablenotemark{2}}                         &
\colhead{${\epsilon}$(S1$_{\rm d}$)}           &
\colhead{${\sigma}1_{\rm d}$\tablenotemark{3}}                  &
\colhead{S2}                                   & 
\colhead{${\epsilon}$(S2)}                     &
\colhead{${\sigma}2$}                          &
\colhead{${\chi}^2$}                           &
\colhead{comments}                             \\
\colhead{}                                     &
\colhead{${\mu}$m}                             &
\colhead{$\frac{\rm MJy}{\rm sr}$}             &
\colhead{$\frac{\rm MJy}{\rm sr}$}             &
\colhead{$\frac{\rm MJy}{\rm sr}$}             &
\colhead{$\frac{\rm MJy}{\rm sr}$}             & 
\colhead{$^{\prime\prime}$}                    &
\colhead{$\frac{\rm MJy}{\rm sr}$}             &
\colhead{$\frac{\rm MJy}{\rm sr}$}             &
\colhead{$^{\prime\prime}$}                    &
\colhead{}                                     \\
}
\startdata
  66 &  60 &  8.33 & 1.35 & 7.60 & 0.91 & 52.40 & 8.55 & 0.76 & 4.63&1.12 &\\
     & 100 & 11.47 & 1.95 & 18.21 & 1.11 & 56.23 & 5.49 & 0.91 & \nodata&2.21 &\\
1554 &  60 & 54.22 & 2.22 & 10.50 & 1.10 & 31.23 & 24.81 &
1.64 & 18.60 & 4.36 & modified beam\tablenotemark{4}\\
     & 100 & 29.57 & 2.87 & 27.61 & 1.51 & 35.04 & 28.59 &
1.72 & 25.29 & 10.21 & modified beam\tablenotemark{4}\\
1727 & 100 & 17.78 & 4.92 & 29.78 & 1.26 & 43.51 & 14.61 &
0.72 &  59.42 & 0.15 & modified beam\tablenotemark{4}\\
\enddata
\tablenotetext{1}{The amplitude S1$_n$ of the nuclear component;}
\tablenotetext{2}{The amplitude S1$_d$ of the disk component;}
\tablenotetext{3}{The width ${\sigma}1_{\rm d}$ of the disk major axis in 
arcsec, where the total width of the major axis of the fitted Gaussian for 
the disk component is 
${\sigma}1_{\rm total}^2={\sigma}1_{\rm d}^2+{\sigma}_{\rm PSF}^2$;}
\tablenotetext{4}{In some cases it was necessary to consider a narrower beam
for the fit, namely a spherical Gaussian with the width
${\sigma}=15.41^{\prime\prime}$, ${\rm FWHM}=2.355 \times {\sigma}$ (see Sect. 6.2.1).}
\end{deluxetable}

\clearpage

\begin{deluxetable}{rrrrrrrrrrrr}
\tablenum{6}
\tabletypesize{\scriptsize}
\tablewidth{400.pt}
\rotate
\tablecaption{The Fitting Parameters for VCC~1673/VCC~1676}
\tablehead{
\multicolumn{1}{c}{} & \multicolumn{5}{l}{VCC~1676} &
\multicolumn{5}{l}{VCC~1673}\\
%\colhead{(1)}                                  &
%\colhead{(2)}                                  &
%\colhead{(3)}                                  &
%\colhead{(4)}                                  &
%\colhead{(5)}                                  &
%\colhead{(6)}                                  &
%\colhead{(7)}                                  &
%\colhead{(8)}                                  &
%\colhead{(9)}                                  &
%\colhead{(10)}                                 &
%\colhead{(11)}                                 &
%\colhead{(12)}                                 \\
\colhead{$\lambda$}                            &
\colhead{S1$_{\rm n}$}                         &
\colhead{${\epsilon}$(S1$_{\rm n}$)}           &
\colhead{S1$_{\rm d}$}                         &
\colhead{${\epsilon}$(S1$_{\rm d}$)}           &
\colhead{${\sigma}1_{\rm d}$}        &
\colhead{S2$_n$}                     & 
\colhead{${\epsilon}$(S2$_n$)}       &
\colhead{S2$_d$}                     &
\colhead{${\epsilon}$(S2$_d$)}       &
\colhead{${\sigma}2_{\rm d}$}        &
\colhead{${\chi}^2$}                \\
\colhead{${\mu}$m}                  &
\colhead{$\frac{\rm MJy}{\rm sr}$}  &
\colhead{$\frac{\rm MJy}{\rm sr}$}  &
\colhead{$\frac{\rm MJy}{\rm sr}$}  &
\colhead{$\frac{\rm MJy}{\rm sr}$}  &
\colhead{$^{\prime\prime}$}         &
\colhead{$\frac{\rm MJy}{\rm sr}$}  &
\colhead{$\frac{\rm MJy}{\rm sr}$}  &
\colhead{$\frac{\rm MJy}{\rm sr}$}  &
\colhead{$\frac{\rm MJy}{\rm sr}$}  &
\colhead{$^{\prime\prime}$}         \\
}
\startdata
170 & 29.45 & 2.72 & 54.52 & 2.03 & 52.68 & 57.08 & 0.63 & 9.18 & 0.10 &
239.00 & 0.2 \\
\enddata
\end{deluxetable}

%% file: tab7.tex
\clearpage

\begin{deluxetable}{rrrrrrrrl}
\tablenum{7}
\tabletypesize{\scriptsize}
\tablewidth{380.pt}
\tablecaption{The Photometry}
\tablehead{
%\colhead{(1)} &
%\colhead{(2)} &
%\colhead{(3)} &
%\colhead{(4)} &
%\colhead{(5)} &
%\colhead{(6)} &
%\colhead{(7)} &
%\colhead{(8)} &
%\colhead{(9)} \\
\colhead{VCC}                               &
\colhead{${\lambda}$}                       &
\colhead{F$_{\rm model}$\tablenotemark{1}}                   &
\colhead{${\epsilon}({\rm F_{\rm model}}$)} &
\colhead{F$_{\rm map}$\tablenotemark{2}}                     &
\colhead{${\epsilon}({\rm F_{\rm map}}$)}   &
\colhead{B$_{\rm backgr}$\tablenotemark{3}}                  &
\colhead{f$_{\rm B}$\tablenotemark{4}}                       &
\colhead{notes}\\
\colhead{}                                  &
\colhead{${\mu}$m}                          &
\colhead{Jy}                                &
\colhead{Jy}                                &
\colhead{Jy}                                &
\colhead{Jy}                                &
\colhead{$\frac{\rm MJy}{\rm sr}$}          &
\colhead{}                                  &
}
\startdata                   
   1&	60 &   0.10 &	0.02 &	-0.05 &	 0.06 &	 15.87 &  0.037 & \\
    &  100 &   0.14 &	0.02 &	 0.21 &	 0.06 &	  8.92 &  0.042 & \\
    &  170 &   0.33 &	0.05 &	 0.31 &	 0.18 &	  7.46 &  0.042 & \\
  17&	60 &$<$0.04 &\nodata &	 0.11 &	 0.08 &	 14.80 &  0.037 & \\	 
    &  100 &$<$0.03 &\nodata &	-0.09 &	 0.08 &	  9.10 &  0.033 & \\
    &  170 &$<$0.07 &\nodata &	 0.01 &	 0.19 &	  7.94 &  0.036 & \\
  24&	60 &$<$0.04 &\nodata &	 0.11 &	 0.06 &	 16.77 &  0.030 & \\
    &  100 &$<$0.04 &\nodata &	-0.08 &	 0.06 &	  9.72 &  0.043 & \\
\enddata
\tablenotetext{1}{The flux F$_{\rm model}$ derived from 
integrating the model fit to infinity. The upper 
limits are indicated with a \lq$<$\rq\ sign in front;}
\tablenotetext{2}{The flux  F$_{\rm map}$ derived from 
integrating the raw maps;}
\tablenotetext{3}{The average background brightness 
B$_{\rm backgr}$;}
\tablenotetext{4}{The $1\,{\sigma}$ fractional uncertainty 
f$_{\rm B}$ in the background brightness, determined from the spread in the 
response of individual detector pixels to the background at the ends of the 
scans.}
\end{deluxetable}